\documentclass[longauth]{aa}  

\usepackage{txfonts}
\usepackage{rotating}
\usepackage{tabularx}
\usepackage{xcolor}
\usepackage{ragged2e}
\usepackage{lipsum}
\usepackage{lscape}
\usepackage{pdflscape}
\usepackage{hyperref}
\usepackage{xcolor}
\usepackage{natbib}

\def\spb{\smallskip\par\noindent$\bullet\;$}
\begin{document}

   \title{COALAS: I. ATCA CO(1-0) survey and luminosity function in the Spiderweb protocluster at $z=2.16$}

   \authorrunning{Jin et al.}

\author{S. Jin\inst{1,2}
  \and H. Dannerbauer\inst{1,2}
  \and B. Emonts\inst{3}
  \and P. Serra\inst{4}
  \and C. D. P. Lagos\inst{5,6,7}
  \and A. P. Thomson\inst{8}
  \and L. Bassini\inst{9,10,11}
  \and M. Lehnert\inst{12}
  \and J. R. Allison\inst{13}
  \and J. B. Champagne\inst{14}
  \and B. Inderm\"uhle\inst{15}
  \and R. P. Norris\inst{16,15}
  \and N. Seymour\inst{17}
  \and R. Shimakawa\inst{18}
  \and C. M. Casey\inst{14}
  \and C. De Breuck\inst{19}
  \and G. Drouart\inst{17}
  \and N. Hatch\inst{20}
  \and T. Kodama\inst{21}
  \and Y. Koyama\inst{18}
  \and P. Macgregor\inst{15}
  \and G. Miley\inst{22}
  \and R. Overzier\inst{23}
  \and J. M. P\'{e}rez-Mart\'{i}nez\inst{21}
 \and J. M. Rodr\'{i}guez-Espinosa\inst{1,2}
  \and H. R\"{o}ttgering\inst{22}
  \and M. S\'anchez Portal\inst{24}
  \and B. Ziegler\inst{25}
          }

   \date{Received 24 December, 2020; accepted 15 March, 2021}

 \abstract
{
We report a detailed CO(1-0) survey of a galaxy protocluster field at $z=2.16$, based on 475 hours of observations with the Australia Telescope Compact Array. We constructed a large mosaic of 13 individual pointings, covering an area of 21 arcmin$^{2}$ and $\pm6500$~km/s range in velocity.
We obtain a robust sample of 46 CO(1-0) detections spanning $z=2.09-2.22$, constituting the largest sample of molecular gas measurements in protoclusters to date. The CO emitters show an overdensity at $z=2.12-2.21$, suggesting a galaxy super-protocluster or a protocluster connected to large-scale filaments with $\sim$120~cMpc size. We find that 90\% CO emitters have distances $>0'.5-4'$ to the center galaxy, indicating that small area surveys would miss the majority of gas reservoirs in similar structures. Half of the CO emitters have velocities larger than escape velocities, which appears gravitationally unbound to the cluster core. These unbound sources are barely found within the $R_{200}$ radius around the center, which is consistent with a picture in which the cluster core is collapsed while outer regions are still in formation. Compared to other protoclusters, this structure contains relatively more CO emitters with relatively narrow line width and high luminosity,  indicating galaxy mergers. We use these CO emitters to place the first constraint on the CO luminosity function and molecular gas density in an overdense environment. The amplitude of the CO luminosity function is 1.6$\pm$0.5 orders of magnitudes higher than observed for field galaxy samples at $z\sim2$, and one order of magnitude higher than predictions for galaxy protoclusters from semi-analytical SHARK models. We derive a high molecular gas density of $0.6-1.3\times10^{9}$~$M_\odot$~cMpc$^{-3}$ for this structure, consistent with predictions for cold gas density of massive structures from hydro-dynamical DIANOGA simulations. 
}


  \institute{
   Instituto de Astrofísica de Canarias (IAC), E-38205 La Laguna, Tenerife, Spain; \email{shuowen.jin@gmail.com}
    \and Universidad de La Laguna, Dpto. Astrofísica, E-38206 La Laguna, Tenerife, Spain
    \and National Radio Astronomy Observatory, 520 Edgemont Road, Charlottesville, VA 22903, USA
    \and INAF - Osservatorio Astronomico di Cagliari, Via della Scienza 5, 09047, Selargius, CA, Italy
    \and International Centre for Radio Astronomy Research (ICRAR), M468, University of Western Australia, 35 Stirling Hwy, Crawley,
WA 6009, Australia
  \and ARC Centre of Excellence for All Sky Astrophysics in 3 Dimensions (ASTRO 3D)
  \and Cosmic Dawn Center (DAWN)
   \and The University of Manchester, Oxford Road, Manchester, M13 9PL, UK
   \and Astronomy Unit, Department of Physics, University of Trieste, via Tiepolo 11, I-34131 Trieste, Italy
   \and INAF - Osservatorio Astronomico di Trieste, via Tiepolo 11, I-34131 Trieste, Italy
   \and IFPU - Institute for Fundamental Physics of the Universe, Via Beirut 2, 34014 Trieste, Italy
   \and Sorbonne Universite\', CNRS, UMR 7095, Institut d'Astrophysique de Paris, 98bis bvd Arago, F-75014 Paris, France
   \and Sub-dept of Astrophysics, Physics, University of Oxford, Denys Wilkinson Building, Keble Road, Oxford OX1 3RH, UK
   \and Department of Astronomy, The University of Texas at Austin, 2515 Speedway Blvd Stop C1400, Austin, TX 78712, USA
   \and CSIRO Astronomy and Space Science, PO Box 76, Epping,
NSW 1710, Australia
   \and Western Sydney University, Locked Bag 1797, Penrith, NSW 2751, Australia
   \and International Centre for Radio Astronomy Research, Curtin
University, Bentley, WA 6102, Australia
   \and Subaru Telescope, National Astronomical Observatory of Japan, National Institutes of Natural Sciences, 650 North A'ohoku Place, Hilo, HI 96720, USA
   \and European Southern Observatory, Karl-Schwarzschild-Stra{\ss}e 2, 85748 Garching, Germany
   \and School of Physics and Astronomy, University of Nottingham, University Park, Nottingham NG7 2RD, UK
   \and Astronomical Institute, Tohoku University, Aoba-ku, Sendai 980-8578, Japan
   \and Leiden Observatory, PO Box 9513, 2300 RA Leiden, The Netherlands
   \and Observat\'orio Nacional, Rua General Jos\'e Cristino, 77, Sao Crist\'ovao, Rio de Janeiro, RJ CEP 20921-400, Brazil
  \and Instituto de Radioastronom\'ia Milim\'etrica, Av. Divina Pastora 7, N\'ucleo Central, E-18012 Granada, Spain
  \and University Vienna, Department of Astrophysics, T\"urkenschanzstra{\ss}e 17, 1180 Wien, Austria
             }

\maketitle

\section{Introduction}
Galaxy protoclusters (see \citealt{Overzier2016} for a review) are expected to contribute significantly to the star formation rate density (SFRD) at high redshifts \citep{Chiang2017cluster}, e.g., 20--50\% at $z>2$. 
Thus understanding how clusters assembled their mass is of critical importance. Molecular gas is the reservoir where the stars form. Measuring the molecular gas content and distribution in galaxies as a function of environment and epoch is crucial for developing our understanding of how galaxies form and evolve. Moreover, since the potential impact of environment can be large on galaxy evolution as evidenced by the tightness of the color-luminosity sequence in local clusters and the morphology-density relation \citep{Dressler1980cluster,Davies2014Virgo}, constraining the evolution of the gas galaxy in clusters/protoclusters offers us key tests of galaxy evolution models.

As a proxy for the cold molecular gas content, CO surveys of random fields \citep[e.g.,][]{Walter2016ASPECS,Riechers2019COLDz,Decarli2019COLF,Decarli2020COLF,Riechers2020COLF} have been revealing the evolution of the molecular gas density over cosmic history. The ASPECS survey \citep{Walter2016ASPECS} conducted the first blind CO search with Atacama Large Millimeter Array (ALMA) in the Hubble Ultra Deep Field (HUDF). This survey revealed an elevated CO luminosity function at $z\sim2$ with respect to the local Universe and an evolution of cosmic molecular gas density within galaxies as a function of redshift \citep{Decarli2016COLF}. Using the Karl G. Jansky Very Large Array (VLA) observations, the COLDz project \citep{Riechers2019COLDz} provided detailed measurements of the CO luminosity function and molecular gas density at $z=2-3$ and $z=5-7$ via CO(1-0) and CO(3-2) detections, respectively. Recently, based on the reported CO detections by \cite{Gonzalez-Lopez2019ASPECS} and \cite{Aravena2019ASPECS}, \cite{Decarli2019COLF,Decarli2020COLF} constructed the CO luminosity function out to $z\sim4$ and found that the observed evolution of the molecular gas density tracks the evolution of the cosmic star formation rate density. This finding is confirmed by the PHIBSS2 survey \citep{Lenkic2020COLF} with CO observations with the Plateau de Bure Interferometer (PdBI). These CO surveys  focus on field galaxy samples while the molecular gas density in dense environments such as galaxy (proto)clusters has not yet been constrained.

Over the last few years, there have been significant efforts to observe CO in high-redshift (proto)cluster environments (e.g., \citealt{Emonts2016Sci,Emonts2018,Hayashi2017CO21_cluster,Noble2017cluster,Noble2019cluster,Casey2016cluster,Wang_T2016cluster,WangTao2018CO,Dannerbauer2017disk,Rudnick2017cluster,Stach2017cluster,Oteo2018cluster,Coogan2018,Miller2018cluster_z4,Tadaki2019cluster,Gomez-Guijarro2019,Lee2019cluster,Castignani2019cluster,Hill2020cluster,Ivison2020cluster,Sperone-Longin2020}). 
However, due to the frequency coverage, ALMA line surveys  mostly target high-J CO($J$, $J-1$) transitions ($J\geq 2$) that could introduce uncertainties into the total molecular gas mass due to unknown gas excitation \citep[e.g.][]{Dannerbauer2009,Daddi2015,Liudz2015,Yajima2020COr21}. The ground-state CO transition (rest-frame 115.27 GHz, $J$=1) is extensively used as the best tracer of the total cold molecular gas mass \citep[e.g.,][]{Ivison2011,Emonts2013}. Using deep CO(1-0) observations with the VLA, \cite{Rudnick2017cluster} revealed two gas-rich cluster members at $z=1.62$ and found that cluster members have comparable gas fractions and star-formation efficiencies (SFEs) with respect to field galaxies, consistent with field scaling relations between the molecular gas content, stellar mass, SFR and redshift. In the $z=2.5$ CL J1001 (proto)cluster field, \cite{Wang_T2016cluster,WangTao2018CO} conducted VLA observations of the CO(1-0) transition. They found low gas content and elevated SFEs in cluster members compared to field galaxies, presenting evidence for impact of the environment on the molecular gas reservoirs and SFEs of $z=2$ cluster galaxies. However, the above mentioned CO(1-0) observations in high-z galaxy clusters targeted either individual cluster members or the central region of a (proto)cluster core (e.g., \citealt{Emonts2016Sci,Emonts2018,Dannerbauer2017disk,WangTao2018CO}; Champagne et al. submitted) and thus introduced a bias in their interpretation in the context of galaxy formation and evolution, as protoclusters can be extended up to 30$^{\prime}$ on the sky \citep[resp. up to 15~Mpc in physical units;][]{Muldrew2015cluster,Casey2016cluster} and superstructure has been found with a size of $\sim120$ cMpc in comoving size at $z>2$ \citep{Cucciati2018}. Clearly, surveys covering the large volume of a galaxy protocluster in the distant universe are still missing and would allow us to characterize the CO luminosity function and molecular gas density in such an environment.

To this end, the Spiderweb protocluster field at $z=2.16$, being a prominent and well-studied example of a cluster in formation, is an ideal laboratory for such a study \citep{Miley2006}. This protocluster has been continually observed and resulting in rich, multi-wavelength datasets. It hosts a significant overdensity of Ly$\alpha$ emitters (LAEs; \citealt{Pentericci2000,Kurk2000}), H$\alpha$ emitters (HAEs) \citep{Kuiper2011,Koyama2013cluster,Shimakawa2014HAE,Shimakawa2018SW}, extremely red objects \citep[ERO;][]{Kurk2004a}, and submilimeter galaxies (SMGs) \citep{Rigby14,Dannerbauer2014LABOCA}. A 10~Mpc scale filament traced by HAEs has been identified across this region \citep{Koyama2013cluster,Shimakawa2018SW}. Meanwhile, the SFRD based on far-infrared/submillimeter measurements appears high, $\sim1500~M_{\odot}$~yr$^{-1}$~Mpc$^{-3}$ \citep{Dannerbauer2014LABOCA}. 
Recently, massive gas reservoirs in this structure have been revealed via CO(1-0) observations with the ATCA. \cite{Emonts2016Sci} found a large reservoir of molecular gas extending across 70 kpc around the central starbursting radio galaxy MRC 1138--262. This reservoir of molecular gas fuels in-situ star formation \citep{Hatch2008Spiderweb} and drives the growth of a massive central-cluster galaxy, suggesting that the brightest cluster galaxies could form out of extended, recycled gas at high redshift.This work demonstrated the power of ATCA to detect the most extended CO(1-0) emission. Subsequently, \cite{Dannerbauer2017disk} discovered an extended rotating disk (40~kpc) with a massive molecular gas mass $M_{\rm mol}$ = 2$\times$10$^{11}$~$M_\odot$ in a normal star-forming galaxy in this protocluster, and found no evidence of environmental impact on the star-formation efficiency (SFE). More recently, one more protocluster galaxy was found in CO(1-0) in the region between the radio galaxy and HAE229 \citep{Emonts2018}. As these CO(1-0) observations target only a small area around the center radio galaxy, the very limited number of detections cannot constrain the CO luminosity function and total gas density in co-moving volume. On the other hand, the physical scale of the protocluster is unknown. This is due to the narrowness of the redshift range probed in narrow-band imaging surveys and the rest-frame UV/optical lines of dusty member galaxies can be severely attenuated, making the identification of cluster membership difficult to complete.
A large scale CO survey with a broad bandwidth will help to find additional cluster members that are missed by optical/near-infrared surveys and could be able to reveal wider structure. Moreover, the SFRD found in \cite{Dannerbauer2014LABOCA} is well above the predictions of simulations and semi-analytical models (e.g., \citealt{Bassini2020simu,Lim2020simu}). It is unclear if the molecular gas density compared to the field sample is similarly overdense as the SFRD, and would be of great interest to compare the observed gas density to predictions from cosmological simulations (e.g., \citealt{Lagos2020,Bassini2020simu}). 
Thus, a dedicated CO(1-0) survey is urgently needed in order to constrain the molecular gas density in this structure and help to clarify these issues.

In the present study, we report the first result of the COALAS (CO ATCA Legacy Archive of Star-forming galaxies) project, focusing on the ATCA CO(1-0) observations of the Spiderweb cluster field at $z=2.16$.
In section~\ref{sec:linesearch}, we discuss the source extraction and construct the CO-emitter catalogue. In section~\ref{sec:molgas}, we present the results of the catalogue. In addition, we constrain the CO(1-0) luminosity function and molecular gas density of this structure. We close the  paper with a discussion in section \ref{sec:discussion} and present the major conclusions of this study in section \ref{sec:conclusion}. 
We adopt a flat $\Lambda$CDM cosmology with $H_{0}=71~$ km s$^{-1}$ Mpc$^{-1}$, $\Omega_{M}=0.27$, $\Lambda_{0}=0.73$ \citep{Spergel2003,Spergel2007}, and a Chabrier IMF \citep{Chabrier2003}.

\section{Observations and data reduction}
\label{sec:observations}

\begin{table*}
{
\centering
\caption{Overview of observations}\label{tab:1}
\label{tab:observations}
\centering
\begin{tabular}{ccccccc}
\hline
\hline
  \multicolumn{1}{c}{Pointing} &
  \multicolumn{1}{c}{Phase center} &
  \multicolumn{1}{c}{Beam} &
  \multicolumn{1}{c}{PA} &
  \multicolumn{1}{c}{Integrated time} &
  \multicolumn{1}{c}{RMS} &
  \multicolumn{1}{c}{Configuration$^a$}\\
 &  &   & [deg] & [hours per config.] & [$b$] &  \\
\hline
 MRC1138   & 11:40:48.3,-26:29:11 & $4.8''\times3.5''$ & 45.5 & 5, 18, 30, 26, 11 &  0.13 &  H75, H168, H214, 750A/D, 1.5A\\
 HAE229   &  11:40:46.6,-26:29:11 & $4.9''\times3.7''$ & -2.7 & 40, 42 & 0.13 &  H214, 750C \\
 DKB01-03  & 11:40:59.2,-26:30:43 & $4.7''\times4.3''$ & 29.6 & 24, 24 & 0.14 & H214, 750D\\
 DKB12 	& 11:40:57.6,-26:29:36	 & $5.1''\times4.3''$ & -86.3 & 23 & 0.20 & H214 \\
 DKB15	& 11:40:54.7,-26:28:21	 & $5.9''\times4.5''$ & 77.0 & 22, 10 & 0.15 &  H214, H168 \\
 DKB16	&	11:41:02.4,-26:27:45 & $5.3''\times4.3''$ & 80.0  & 18 & 0.23  &  H214 \\
 SWpoint1 & 11:40:37.7,-26:30:20 & $9.8''\times7.6''$ & 69.0 & 15, 16 &  0.18 &   H168, H75\\
 SWpoint2 & 11:40:43.1,-26:29:20 & $6.5''\times5.9''$ & 66.0 & 15, 5 &  0.18 &  H168, H75 \\
 SWpoint3 & 11:40:44.1,-26:28:32 & $7.0''\times6.2''$ & 74.0 & 20, 6  & 0.19  &  H168, H75 \\
 SWpoint4 & 11:40:52.5,-26:29:30 & $13.8''\times10.7''$  & -84.9 & 8, 22 &  0.24 & H168, H75 \\
 SWpoint5 & 11:41:00.1,-26:28:56 & $12.6''\times9.8''$ & -83.9   & 9, 20 &  0.22 & H168, H75  \\
 SWpoint6 & 11:40:40.1,-26:29:47 & $13.8''\times13.1''$ &  -87.0 & 27 & 0.19  &  H75 \\
 SWpoint7 & 11:40:39.7,-26:28:45 & $12.7''\times9.8''$  & -86.7  & 8, 15 & 0.29 &  H168, H75 \\
  \hline\hline\end{tabular}\\}
\tablefoot{
$^a$ See \href{https://www.narrabri.atnf.csiro.au/operations/array_configurations/configurations.html}{www.narrabri.atnf.csiro.au/operations/array\_configurations/configurations.html} for details on the array configurations.

$^b$  In unit of mJy~${\rm bm^{-1}}$ in 40~km~s$^{-1}$ channel width.
}
\end{table*}

The CO ATCA Legacy Archive of Star-Forming Galaxies (COALAS) project\footnote{http://research.iac.es/proyecto/COALAS/pages/en/home.php} is a large program (ID: C3181, PI: H. Dannerbauer) with the Australia Telescope Compact Array (ATCA). The observations took place during April 2017 until March 2020. A major goal of this project is to study the impact of environment on the cold molecular gas content via the CO(1-0) transition. In total, we obtained 820~hours observing time with the ATCA (including compensation for bad weather). {The COALAS large program} consists of well-covered `field'-targets from the {Extended Chandra Deep Field-South (ECDFS)} and `cluster'-targets from the Spiderweb protocluster ($z=2.16$). 
In this work, we present only the results in the Spiderweb protocluster field, while the work in ECDFS will be presented in Thomson et al. (in prep.).

In addition, we included observations of several LABOCA-selected sources (ID: 2014OCTS/C3003 and 2016AOPRS/C3003: PI: H. Dannerbauer, high resolution pointings focusing on DKB01-03 ID: 2017APRS/C3003, PI: B. Emonts), the center radio galaxy and the HAE229 \citep{Emonts2016Sci,Dannerbauer2017disk}. {These observations were finished before the large program started, which were conducted in the same Spiderweb field at the same frequency as the large program.}
In total, we have observed the Spiderweb protocluster field for 475~hours (including bad weather).
We stress that currently, ATCA is the only facility in the southern hemisphere that can target the lowest CO transition of $z\sim2$ galaxies. The situation will change with the installation of ALMA band~1. 

As shown in Fig.~\ref{fig:snmap}, including data from \cite{Emonts2016Sci,Emonts2018} and \cite{Dannerbauer2017disk}, 13 pointings in total have been observed at 36.5 GHz (the redshifted CO(1-0) transition) with a 2~GHz bandwidth. 
The pointings have been designed in that way to maximize the number of known sources (with spectroscopic redshifts from the rest-frame UV/optical) per pointing. 
Therefore, the spacing between the pointing varies, and the depth of the mosaic is not homogeneous. The pointings {cover primarily submillimeter galaxies detected with LABOCA that are most probably physically related to the protocluster \citep{Dannerbauer2014LABOCA}, secondarily the HAEs confirmed in the same structure  \citep{Koyama2013cluster,Shimakawa2014HAE}}. We observed at different configurations for about 8~hours per night, mostly using the most compact array configuration as H75, H168 or H214 (see details in Table~\ref{tab:observations}), and the beam size varies between 4$^{\prime\prime}$ to 14$''$. Therefore, in principle we do not expect to  resolve  sources in the data cube. However, because of the very compact array configurations (baselines $<100$~m), our ATCA observations are very sensitive to  low-surface-brightness emission from extended molecular gas reservoirs, ensuring that we obtain reliable mass estimates. We summarize the observations of the 13 individual pointings in Table~\ref{tab:observations}, in which 11 of them are from the COALAS large program. The data of the pointings of MRC1138 and HAE229 are taken from previous studies of \cite{Emonts2016Sci,Emonts2018} and \cite{Dannerbauer2017disk}. For the remaining 11 pointings, the phases are calibrated every 15 minutes with a short ($\sim$ 2 min) scan on a nearby calibrator 1124$-$186. 1124$-$186 is also used for bandpass calibration which calibrated out a uniform sensitivity over the frequency range covered. Flux calibration was done on PKS 1934$-$638. Following \cite{Emonts2014}, a conservative $20\%$ uncertainty in the measured fluxes is assumed to account for the uncertainty in absolute flux calibration. However, this does not affect the source extraction via signal-to-noise-ratio (SNR) in the data cube, see section~\ref{sec:linesearch}. 

We use the software packages MIRIAD \citep{Sault1995Miriad} and Karma \citep{Gooch1996Karma} for the data reduction and visualization, following the strategy described in detail in \cite{Emonts2014,Emonts2016Sci} and \cite{Dannerbauer2017disk}. We imaged the ATCA data cube with natural weighting using the task INVERT for each individual pointing. The FWHM of the ATCA primary beam is $70''$ at 7mm. We imaged each individual pointing with a radius of $0.9'$ so that adjacent pointings significantly overlap. The total sky coverage of the 13 ATCA pointings is 25.8 arcmin$^2$, while sources are only detected in regions with rms noise $<$1 mJy per beam which corresponds to an area of 21.3 arcmin$^2$. The data cubes are imaged using pixel size of 1.5$^{\prime\prime}$ and channel width of 40 km s$^{-1}$. We applied a Hanning smooth, resulting in a resolution of 80 km s$^{-1}$. The continuum was separated from the line by fitting a straight line to the line-free channels in the uv-domain. Velocities are in the optical frame relative to $z=2.1612$, spanning $\pm7000$ km/s.

We linearly combined the 13 pointings using the task LINMOS after correcting for primary beam attenuation. The rms level of individual pointing are $0.13-0.29$~mJy per beam (in 40~km~s$^{-1}$ channel) on the phase center of each pointing as listed in Table~\ref{tab:observations}, while the SNR in the overlapping regions got improved, with an rms of $0.1-1$~mJy over the whole field after primary beam correction. The deepest area has an rms of $0.09$~mJy per beam, and is centered on the radio galaxy MRC1138-262 and HAE229 that are covered by $3-4$ individual pointings.  Because none of the new CO signals in our survey is strong enough to have significant sidelobes,  we did not deconvolve our  images, and therefore performed our analysis on the `dirty' cube, following the previous studies in \cite{Emonts2016Sci,Emonts2018} and \cite{Dannerbauer2017disk}.

\section{Line search}
\label{sec:linesearch}
We adopted two major methods to search for CO lines in the ATCA data cube: (i) our code MaxFinder, which is an algorithm searching for the maximum SNR of candidate lines on each pixel and matching counterparts in optical images (Sec.\ref{sub:maxfinder}); and (ii) the Source Finding Application (SoFiA) which is a blind source finding pipeline intended to find and parametrize lines in data cubes \citep{Serra2015SoFiA}, see details in Sec.~\ref{sub:sofia}. As a complementary approach, we search for CO lines using the reshifts of the HAEs in this field.
{We summarize the selection of the final catalogue in Sec.~\ref{subsec:catalogue}. The statistical properties of the catalogue (e.g., completeness and flux boosting) will be described in
  Sec.~\ref{subsec:fluxboosting}.}

\subsection{MaxFinder}
\label{sub:maxfinder}
\begin{figure*}
\centering
\includegraphics[width=0.99\textwidth]{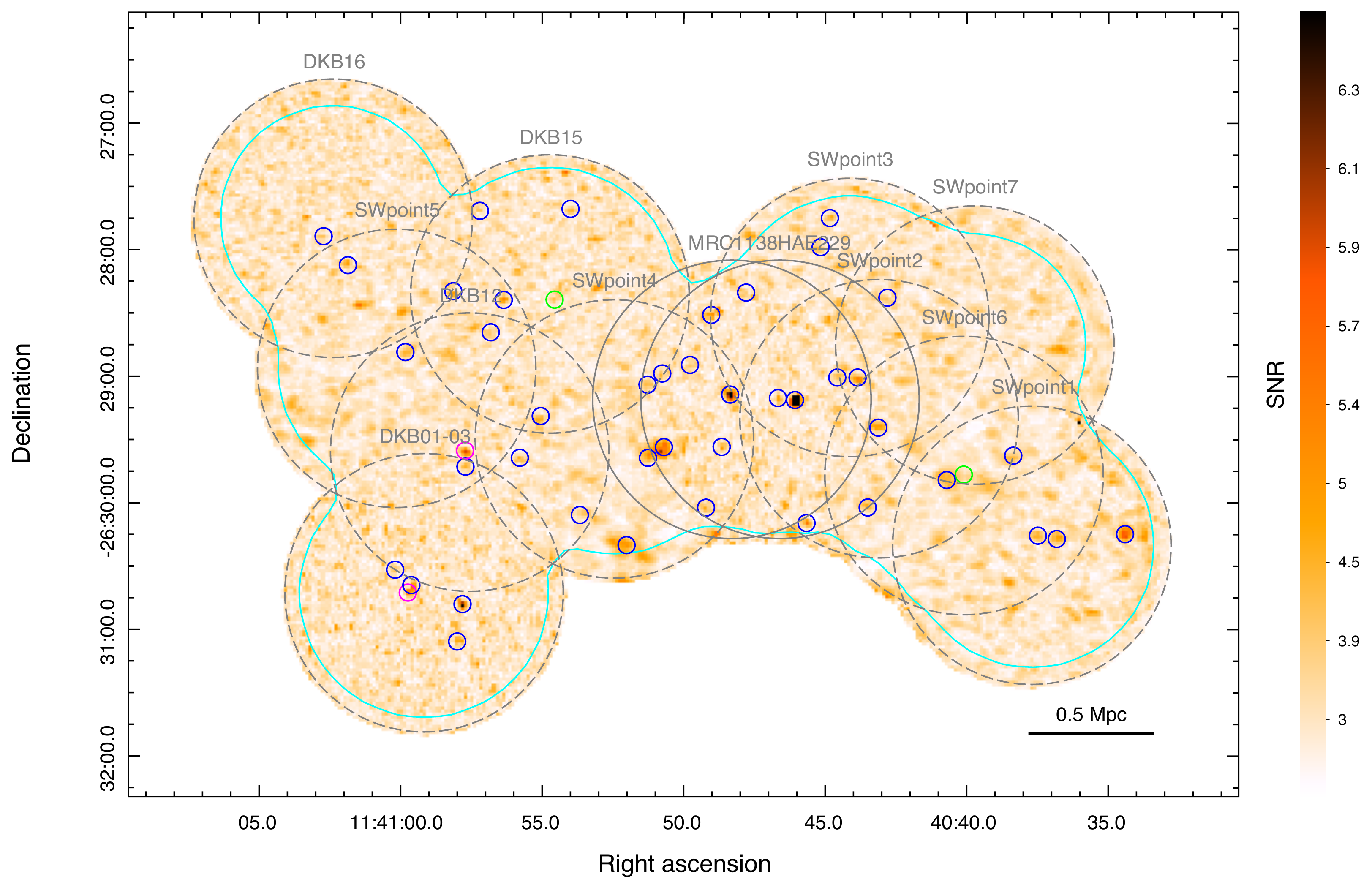}
\caption{
The SNR map of line candidates (Sec.~\ref{sub:maxfinder}) in the ATCA mosaic. The SNR is the maximum signal-to-noise ratio computed in the 1D spectrum on each pixel by our algorithm MaxFinder (see details in Sec.~\ref{sub:maxfinder}). {The color bar values are limited in $1.5<$SNR$<6.5$ for fair visibility of detections. ATCA pointings are shown in grey circles with a radius of $0.9'$. Grey solid circles mark the pointings MRC1138 in \cite{Emonts2016Sci} and HAE229 in \cite{Dannerbauer2017disk}, respectively. Cyan contour encloses the area with rms $<1$~mJy~bm$^{-1}$.} Detected sources are marked by solid circles in colors, blue: category A; magenta: category B; green: category C (see details in Sec.~\ref{subsec:catalogue}). 
 \label{fig:snmap}
}
\end{figure*}

The MaxFinder code is inspired by previous work on manual extraction of CO(1-0) emission from ATCA cubes \citep{Emonts2014,Emonts2016Sci,Dannerbauer2017disk,Emonts2018}. Equivalently, the MaxFinder source extraction code will  search for the most probable line candidate on every position (i.e., pixel) and look for their optical/near-infrared counterparts afterwards. We run the MaxFinder on the large mosaic made of all 13 pointings. This large mosaic contains the best SNR in regions overlapped by multiple pointings albeit with variation in resolution, enabling us to search for the most probable line candidates in the data cube. First, MaxFinder extracts the spectrum on each pixel and calculates the weighted average over any two channels. Given that the sensitivity is uniform over the full ATCA spectral range, the flux uncertainty is defined as the standard deviation of the full spectrum without Hanning smoothing. We compute the weighted average over each two arbitrary channels, and select the two channels $n_{i}$ and $n_j$ that maximize the weighted average. Second, we define the range of line as channels lying in the FWZI (full width at zero intensity) that contains the channels $n_{i}$ and $n_j$. The MaxFinder will determine the FWZI range by searching for the first negative channel on the left (right) side of channel $n_i$ ($n_j$).
Adopting Eq. 2 in \cite{Emonts2014}, the line flux is integrated over the FWZI range, and uncertainty of the line flux is calculated as $\Delta{I_{\rm co}}=\sigma\Delta v\sqrt{{\rm FWZI}/\Delta v}$. Thus the SNR is 
\begin{equation}
    {\rm SNR} = \frac{\int_{\rm FWZI}Sd v}{\sigma\Delta v\sqrt{{\rm FWZI}/\Delta v}}
\end{equation}
where $\sigma$ is the root-mean-square (rms) noise, $\Delta v$ is the channel width and FWZI the range over which the line flux was integrated. In this way we keep the calculations consistent with previous work by our team. Then we calculate signal and noise of candidate line on each pixel and make a large mosaic SNR mosaic, as shown in Fig~\ref{fig:snmap}. Note that the radio galaxy MRC~1138-262 \citep{Emonts2013,Emonts2016Sci} and the HAE229 \citep{Emonts2013,Dannerbauer2017disk} are solidly detected in the SNR mosaic where the peak-SNR on their positions are SNR$_{\rm peak}=$7.5 and 8.1, respectively. This verifies the robustness of our  search method. At last, we select all sources with  SNR$_{\rm peak}>$5 in the large mosaic. For candidates with SNR$_{\rm peak}$=4--5, which have a high possibility of being spurious, we require a counterpart in either the HST F814W (i.e., the $I$ band, \citealt{Miley2006}) or the VLT/HAWK-I K$_{\rm s}$ images (\citealt{Dannerbauer2014LABOCA}). Such approach was already applied on ALMA data to exclude spurious sources \citep[see e.g.,][]{Dunlop2017}. We searched for counterparts within a circle centering on the pixel with peak SNR and with a radius of 3$''$ (i.e., 2 pixels). In order to avoid bad channels on the edge of the sideband, we take only the spectral data within velocity range $\pm6500$ km/s. In this way, we select 75 SNR$>$4 sources with optical counterparts.

\subsection{SoFiA}
\label{sub:sofia}
In order to obtain a secure catalogue, we also applied the blind source extraction method SoFiA \citep{Serra2015SoFiA}\footnote{\href{https://www.atnf.csiro.au/people/Tobias.Westmeier/tools_software_sofia.php}{https://www.atnf.csiro.au/people/Tobias.Westmeier/tools\_software\\ \_sofia.php}} on the ATCA data cubes. SoFiA is a flexible source finder tool for 3D spectral line data. SoFiA is designed for application on any data cube, which is independent of the type of emission line or  telescope. SoFiA allows the user to search for spectral line signal on multiple scales on the sky and in frequency, and it takes into account noise level variations across the data cube and the presence of errors and artefacts which are crucial to detect and parametrize 3D sources in a complete and reliable way. Finally, SoFiA is able to estimate the reliability of individual detection. To quantify the confidence of line detection, SoFiA defines a parameter {\it reliability}
\begin{equation}
    R = \frac{T}{T+F}
\end{equation}
where $T$ and $F$ are the number of true and false detections, respectively, and are estimated based on the combined statistics of sources with positive and negative total flux \citep{Serra2012reliability}.
Thus the $reliability$ is identical to the {\it fidelity} parameter defined in \cite{Gonzalez-Lopez2019ASPECS} (Eq. 1) and the {\it purity} parameter defined in \cite{Bethermin2020ALPINE} (Eq. 1). 
In this study, we run SoFiA with threshold of ${\rm SNR=1.15}$ to generate sufficient negative detections  to constrain the {\it reliability}. Given that the 13 ATCA pointings are observed with different configurations, the large mosaic --- with varying/inconsistent spatial resolution --- 
 is not suitable for blind source finder algorithm like SoFiA. Therefore, we produced four mosaics via linearly combining pointings with comparable beam sizes using LINMOS, i.e., MRC1138$+$HAE229, DKB12$+$15$+$16, SWpoint2$+$3, SWpoint4$+$5 and SWpoint6$+$7. As DKB01$-$03 and SWpoint1 have different spatial resolutions compared to their neighbouring pointings, SoFiA is applied individually on them. The same SoFiA recipe is applied on above mosaics and individual pointings with tailored $reliability.f_{min}$ \footnote{See SoFiA control parameter at https://github.com/SoFiA-Admin/SoFiA/wiki/SoFiA-Control-Parameters} to their beam sizes. This parameter is defined as $reliability.f_{min}={\rm SNR_{t}} \sqrt{A_{beam}}$, where the ${\rm SNR_t}$ is the SNR threshold and $A_{beam}$ is the area of the beam in unit of pixel. It is scaled with the square root of the beam area, ensuring that the integrated SNR threshold is identical for all cubes. On the other hand, as detections are not likely resolved in the ATCA beam, we only smooth the spectra in velocity space, and require {\it reliability}$>$0.5 and line width of at least 3 channels (i.e., $>$120 km/s).
In this way, we detect 187 sources with SoFiA, of which 74 have SNR$_{\rm peak}>4$ within their beams in the SNR map generated by MaxFinder.

\subsection{Source Catalogue}
\label{subsec:catalogue}
The above two methods have both pros and cons. The blind line search generally suffers from a high spurious rate, while searching with priors appears to be more reliable but could still include spurious sources. In detail, although MaxFinder is prior-based, a weakness is that some detections could be spuriously boosted by noise and counterparts can be found by chance on position of the noise peak as MaxFinder is searching for signal on individual pixels and the ATCA beam sizes are 3--8 times larger than the pixel size. 
However, such spurious detections can be filtered out by SoFiA because it is searching for signal in an area of the beam rather than individual pixels, so that spurious sources with a strong negative peak nearby will be excluded from the detection list. 

In order to create a catalogue of reliable line detections, we  combine the two methods, preferentially selecting  sources extracted by both the MaxFinder and SoFiA. We match MaxFinder and SoFiA sources according to spatial positions and line velocities, where the tolerance of spatial offset is half of the beam size in SoFiA mosaic and the velocity difference is less than the width of two FWZIs. 
We find 42 sources detected by both the MaxFinder and SoFiA having a SNR$>4$ (MaxFinder) and $reliability>0.5$ (SoFiA). We classify these sources as category A. In category A, all known CO(1-0) detections of this protocluster from the literature such as MRC1138$-$262 \citep{Emonts2013,Emonts2016Sci}, HAE229 \citep{Emonts2013,Dannerbauer2017disk} and an additional detection by \cite{Emonts2018} are re-discovered by our source extraction, giving credibility for our method. The line intensities of these detections agree well with the results in \cite{Emonts2018}. We obtain higher SNRs for MRC1138$-$262 and HAE229 than the detections presented previously in \cite{Emonts2018} as they used two pointings, whereas in this work we add more visibilities to these sources from overlapped pointings SWpoint2, 3 and 4. On the other hand, these two galaxies are spatially resolved \citep{Emonts2016Sci,Dannerbauer2017disk} and we adopted a large pixel size ($1''.5$) in imaging which also slightly boosted the SNRs. 

The majority of SoFiA sources (SNR$>3.5$) are not included in category A, either because of low SNR ($<$4) or because they lacked an optical/NIR counterpart. 
On the other hand, the SoFiA line search is conducted in mosaics of data with comparable resolutions, i.e., MRC1138$+$HAE229, DKB12$+$15$+$16, SWpoint2$+$3, SWpoint4$+$5 and SWpoint6$+$7 which have lower SNR in overlapping areas than that in the large mosaic of 13 pointings. Therefore, compared to MaxFinder which was applied on the fully combined mosaic, SoFiA could miss detections in overlapping area due to higher noise level. Thus we include sources that are solidly detected by MaxFinder but missed by SoFiA. We find that two sources selected in MaxFinder with SNR$>5$ are missed by SoFiA. One source is the LABOCA-detected SMG DKB12 \citep{Dannerbauer2014LABOCA}. SoFiA missed this source due to a negative noise peak near the source in the low SNR mosaic. The second source is an HAE with a possible counterpart of the LABOCA-selected SMG DKB01a \citep{Dannerbauer2014LABOCA}. The reason that SoFiA missed this source could be that it is blended with the nearby source DKB01b \citep{Dannerbauer2014LABOCA}. We classify the two sources as category B. 

In addition, we also search for CO detections based on HAE priors from \cite{Koyama2013cluster} and \cite{Shimakawa2018SW}. This approach re-discovers 12 HAEs with SNR$>$4 that have been included in category A and B (Tables~\ref{tab:atcasnr5}) which further credit to our search methods MaxFinder and SoFiA. Moreover, we find two CO detections on HAE positions that are not included in category A and B. The two sources are detected with 3.8$\sigma$ and 4.4$\sigma$ respectively while are not selected by SoFiA. However, their CO spectroscopic redshifts are consistent with H$\alpha$ spectroscopy \citep{Shimakawa2018SW}, hence they are securely detected and we classify them to category C. To summarize, we show the definition of categories in Table~\ref{tab:category}. In total, 46 reliable CO detections are included in this catalogue, as listed in Table~\ref{tab:atcasnr5} and shown in Fig.~\ref{fig:snmap}. We note that other SNR peaks in Fig.~\ref{fig:snmap} are not identified as detections, because they lack a reliable counterpart or are not selected by SoFiA.

\begin{table}
{
\centering
\caption{Detection category}\label{t2}
\label{tab:category}
\centering
 \begin{tabular}{|c|c|c|}
\hline
  \multicolumn{1}{|c|}{Category} &
  \multicolumn{1}{c|}{Number} &
  \multicolumn{1}{c|}{Criteria} \\
\hline
  A & 42 &  Detected by SoFiA \& MaxFinder\\
  \hline
  B & 2 & MaxFinder SNR>5 \& \\
     &     &  not detected by SoFiA\\
  \hline
  C & 2 & HAEs with matched $z_{spec}$, \\
       &    &  neither in category A nor B \\
  \hline
\end{tabular}
}
 \end{table}

The MaxFinder and HAE prior-based extraction use 3$^{\prime\prime}$ tolerance for all pointings when matching counterparts, being appropriate for counterpart search in high-resolution pointings (e.g., MRC1138 and HAE229) but could be too strict for low resolution pointings. However, using a larger search beam would introduce high spurious rate as the deep HST I band image contains a large amount of faint sources that could be associated to spurious detections by chance. As a trade off, we decide to apply the 3$^{\prime\prime}$ tolerance for all pointings to obtain a conservative detection list, albeit some detections in low resolution pointings could be missed. We refer to a detailed analysis on false positive rates within the search radius in Sec.~\ref{subsec:fluxboosting}. We also note that the recently found optically-dark dusty galaxies \citep{Jin2019alma,Wang2019Natur,Smail2020} would be not included in this catalogue as we conservatively request a optical counterpart for each detection.

\subsection{Flux boosting, completeness and false positive rate}
\label{subsec:fluxboosting}
To constrain the CO luminosity function in this protocluster. it is important to measure the  completeness of the line detection, and the effects of flux boosting. We therefore performed Monte Carlo simulations in the ATCA large mosaic, analogous to the methods applied in COLDz \citep{Pavesi2018COLDz}, ASPECS \citep{Decarli2019COLF} and PHiBSS2 \citep{Lenkic2020COLF} projects. Given that detections are not expected to be resolved in the ATCA beam, we assume that the probability of line detection (i.e., completeness) only depends on the integrated line flux, on the line width and on the depth of the data cube (i.e., rms level). As the MaxFinder algorithm works on the spectrum of individual pixels, we perform simulations on 1D spectra extracted on randomly selected positions. We inject mock lines in the data cube, spanning a range of values for various parameters (line flux, FWHM and rms level), then search for lines via the MaxFinder algorithm and check the fraction of lines recovered. In detail, the mock line with a Gaussian profile (integrated line flux, FWHM) is co-added to the 1D spectrum on a random pixel selected in an area with given rms value. To cover the parameter space of our line detections and variable depths of ATCA mosaic, we designed 11 rms bins spanning $0.08-1$~myJy/beam, 20 injected line fluxes evenly spanning $0.03-1$~Jy km/s in log space and 16 FWHMs spanning 125--1200km/s in linear space. In each rms bin, we co-added a mock line (in fixed line flux and FWHM) to a 1D spectrum for 100 times where each time the 1D spectrum is extracted on a random position that has an rms value within the rms bin and no line detection on the position. Afterwards, we run MaxFinder on the co-added spectrum to search for lines with maximum SNR, and identify the lines as recovered lines if matching the two criteria: (1) the central velocity of  the output line has a distance from the input line center less than the injected FWHM, and (2) the output line has SNR$>4$.

\begin{figure}
\centering
\includegraphics[width=0.48\textwidth]{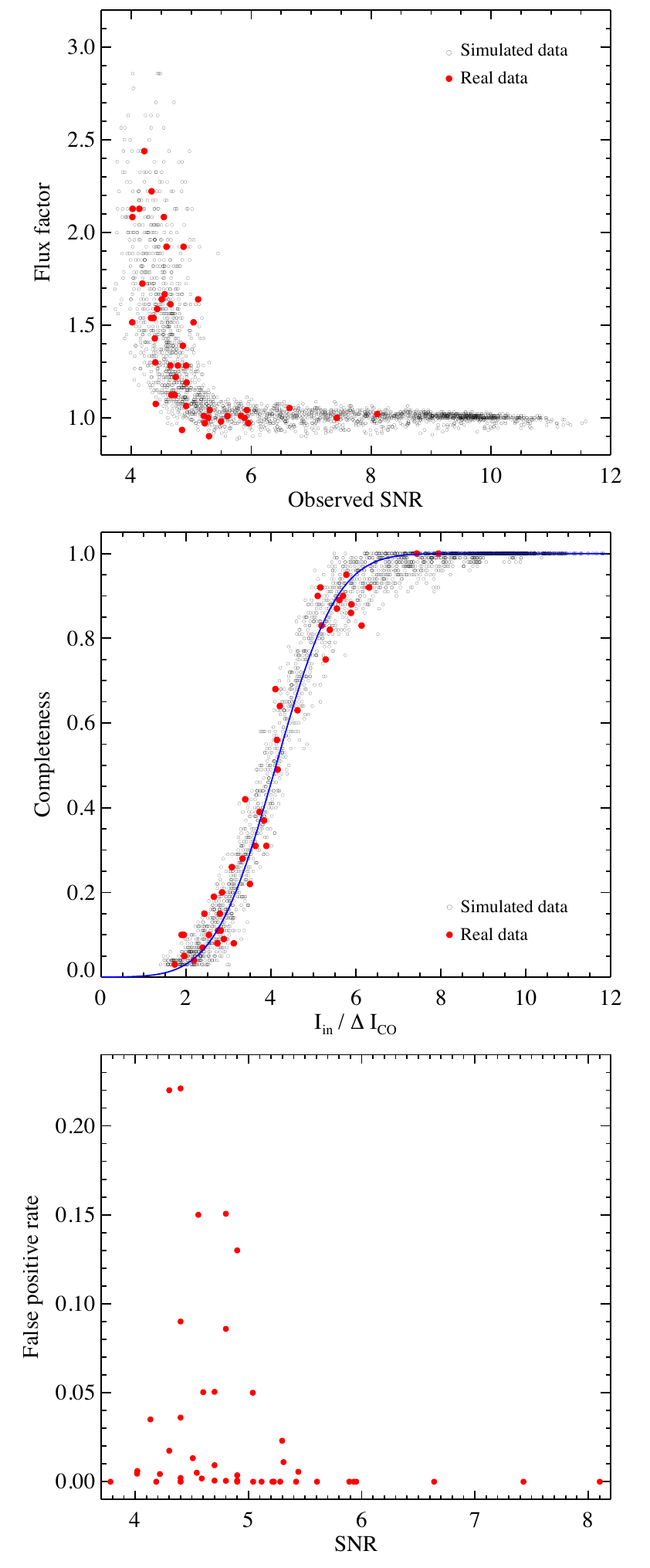}
\caption{%
{\it Top panel:} The flux boosting factor vs. observed SNR for simulated data (open circle) and real detections (red filled circle). 
{\it Central panel:}
The completeness as a function of the injected line flux $I_{\rm in}$ over the measured flux uncertainty $\Delta I_{\rm CO}$. The $I_{\rm in}$ for real detections are line fluxes that have been corrected for flux boosting factors in the top panel. The blue curve shows the best fit to the simulated data (see Eq. 3). The completeness decreases with decreasing integrated flux normalized by noise, indicating that sources are more difficult to detect at low SNR. This analysis allows us to correct CO luminosity functions for the completeness of our search algorithm. 
{\it Bottom panel:} The false positive rate vs. SNR for CO detections in this study.
\label{completeness}
}
\end{figure}

The flux factor $f_{\rm flux}$ is defined as the ratio of the median of the recovered line fluxes to the injected ones, and the completeness $C$ is inferred as the ratio between the number of recovered and injected lines. Simulations are performed 3200 times with varying the values of injected line properties, enabling us to associate the measured line properties to their intrinsic values for all lines that can be detected in the ATCA mosaic. For each detected candidate, based on the measured velocity-integrated intensity $I_{\rm CO}$, line width FWHM and the rms level $\sigma$ on its position, we obtained the flux factor and completeness $C$ via matching the line properties to the recovered ones in the simulations, thus associating to a discrete grid of intrinsic properties. 
{We note that the radial variations of depths caused by the primary beam corrections and different integration have  accounted for the completeness, as different rms levels (depths) have been fully mimicked in the simulations.}
The flux boosting factors in this detection catalogue are spanning $0.9-2.4$ with a median of 1.2. The simulations also provide us with the deviation of the recovered flux density $5-28\%$ with a median of 20$\%$ which is in excellent agreement with the 20$\%$ uncertainty that have been assumed to the observed line fluxes in order to account for uncertainty in flux calibration \citep{Emonts2014}. The completeness is tightly related to the line properties as a function of $I_{\rm in}/\Delta{I_{\rm CO}}$, as shown in Fig.~\ref{completeness}-central where $I_{\rm in}$ is the injected line flux and the $\Delta{I_{\rm CO}}$is the measured flux uncertainty. This relation has a similar expression to the Eq.5 in \cite{Lenkic2020COLF} which is 
\begin{equation}
    C(x) = \frac{1}{2}[1+{\rm erf}(\frac{x-A}{B})]
\end{equation}where $x=I_{\rm in}/\Delta{I_{\rm CO}}$. The best fit shows A$=4.1$ and B$=1.6$ in this study, which is shown in blue curve in Fig.~\ref{completeness}. Given this tight correlation, we assign completeness and flux factor for each detection by matching to a simulated source with the closest SNR, line flux and rms level which are shown in Table~\ref{tab:atcasnr5}. In the following analysis, we scale down the observed CO luminosities in all figures by the flux factors to account for flux boosting while still presenting flux factors and the observed flux and luminosities in tables. We do not scale down the uncertainties in the flux and luminosity to have more conservative results. 

Given that the simulations are computed for secure sources with a known position, this appears to be appropriate for the category A and B sources in our catalogue. Note that one source in category C is detected with SNR$=$3.8 that is lower than the SNR$>$4 threshold applied in the simulations, we thus do not have constraints on its flux factor and completeness. We verified that including this source or not does not impact on the following statistical analysis.
Thus we adopt the completeness and flux factors for category A \& B sources and one source with SNR$>$4 in category C. 

In Fig.~\ref{completeness}-bottom, we show the false positive rate of this sample. Among the 46 CO detections (Table~3), nine sources have well-matched optical spectroscopical redshifts to their CO redshifts which are solidly confirmed and false detection case can be excluded. The rest of the sources have no optical redshifts known. Some of them could be false positives, and counterparts can be found by chance within the $3''$ searching radius. The probability of false line detection can be conservatively estimated as $1-R_{\rm SoFiA}$, where the $R_{\rm SoFiA}$ is the $Reliability$ computed from SoFiA. As the $R_{\rm SoFiA}$ is computed from mosaics with similar resolutions, shallower than the large mosaic, the $Reliability$ of detection is actually higher than the SoFiA output and the probability of false line detection  $(1-R_{\rm SoFiA})$ appears conservative. To constrain the probability of false counterparts, $P_c$, we run simulations via randomly generating mocked sources with the same sky density of HST I and VLT Ks sources in this field, and searching for counterparts using a $3''$ radius on a random position. We find that the false positive rate of HST I band counterparts is $P_c=0.46$, and false Ks counterpart is $P_c=0.06$. The false positive rate of HAE counterparts is entirely negligible, as the HAE sky density is lower than the HST I source density by two orders of magnitude, thus we assigned the false rate of HAE counterparts to 0. Therefore, the false positive rate $P_{\rm false}$, i.e., the probability of a counterpart found by chance for a false line detection, is $P_c\times(1-R_{\rm SoFiA})$. As shown in Fig.~\ref{completeness}-bottom, all sources in this sample have false positive rate $<0.24$, and only sources with SNR$<5.5$ have $P_{\rm false}$>0.  {The total false detection number (i.e., $\Sigma P_{\rm false, i}$) in this sample is less than 1.}

\begin{figure*}
\centering
\includegraphics[width=0.98\textwidth]{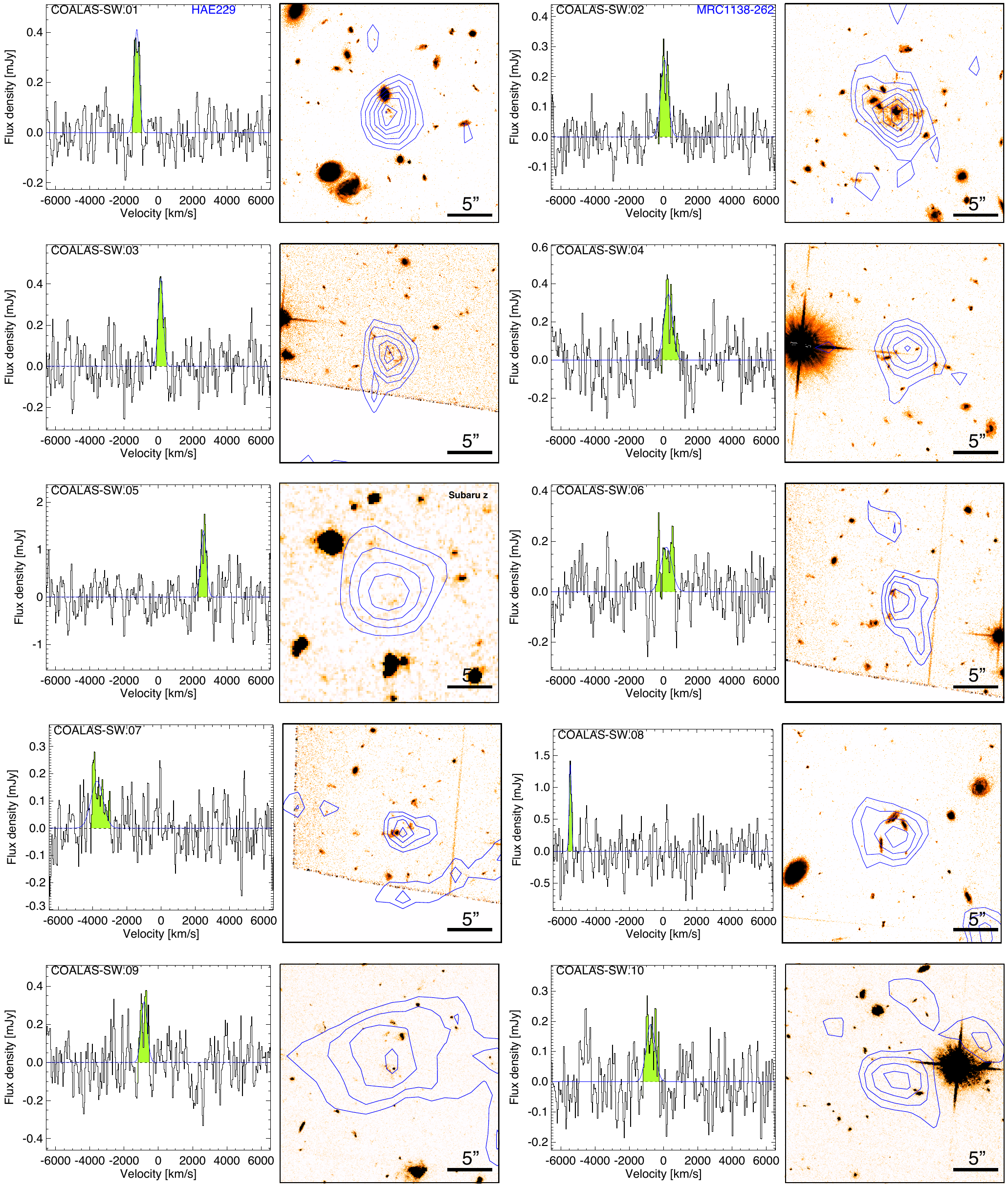}
\caption{
The ATCA spectra and integrated CO maps for detections with SNR$>$5.  {\it Spectra panels:} CO(1-0) lines are highlighted in green and fitted by a single Gaussian profile shown in red curve. 
Reference names in \cite{Emonts2016Sci} and \cite{Dannerbauer2017disk} are highlighted in blue when available. 
{\it Image panels:} Optical images in $25''\times 25''$ size overlaid with CO(1-0) intensity contours. In general, images are taken from the HST F814W data \citep{Miley2006}, while VLT $K_s$ \citep{Dannerbauer2014LABOCA} and Subaru {z-band} images \citep{Koyama2013cluster} are indicated as text on upper-right corner. Contours start at $2\sigma$ in steps of $1\sigma$. 
 \label{fig:spec_sn5_1}
}
\end{figure*}

\addtocounter{figure}{-1}
\begin{figure*}
\centering
\includegraphics[width=0.98\textwidth]{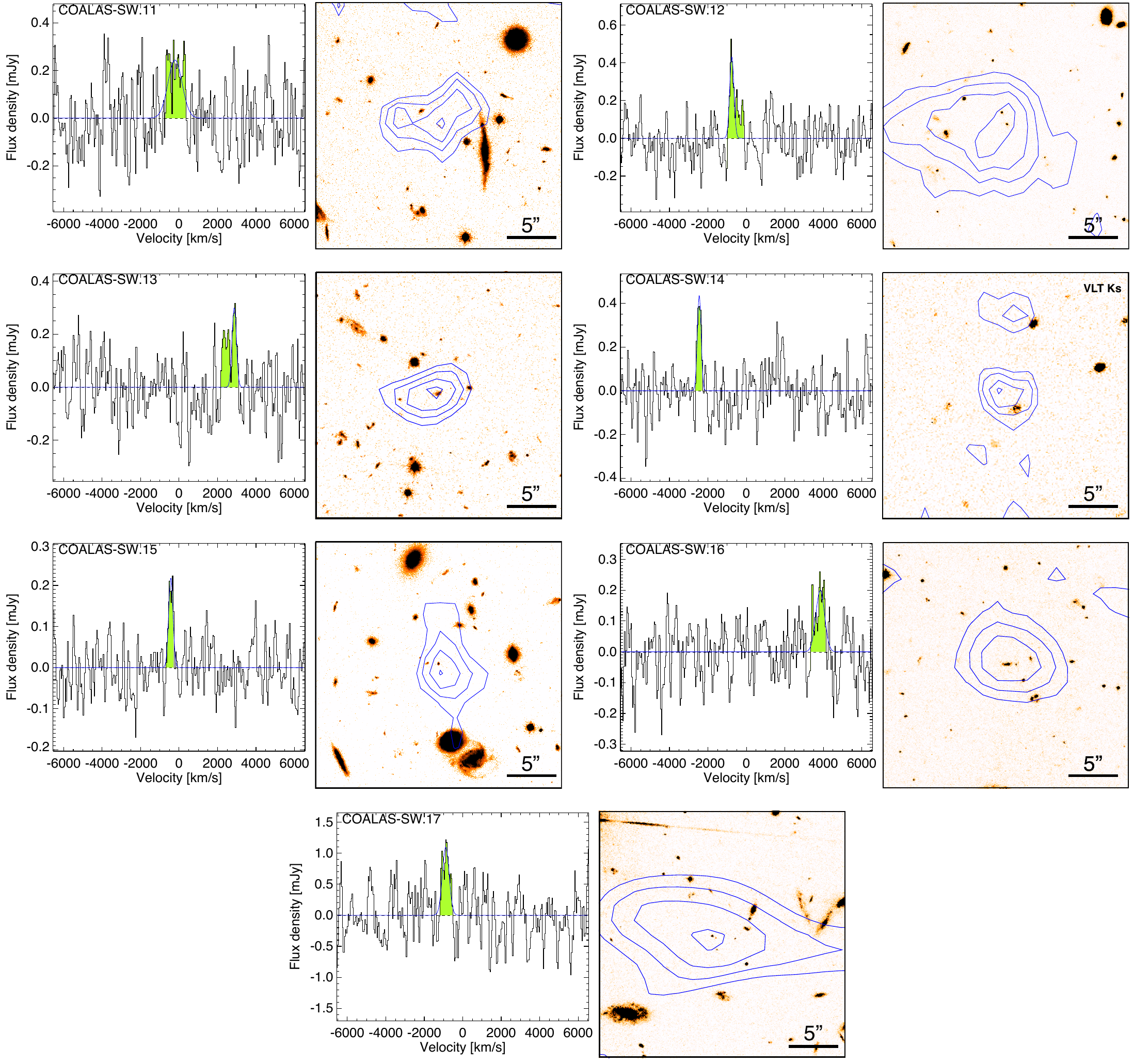}
\caption{
Continued.
 \label{fig:spec_sn5_2}
}
\end{figure*}

\section{Molecular gas content of the Spiderweb protocluster}
\label{sec:molgas}
Based on the source catalogue presented in section~\ref{subsec:catalogue}, we show the individual spectrum for each detection. Furthermore, we study the cold molecular gas properties of our sample including the determination of the CO(1-0) luminosity function of this high-density field.

\subsection{Spectra and integrated CO maps}
\label{sub:spectra}
For all sources, we present the 1D-spectra and coutours of the integrated CO maps overlaid on optical/near-infrared images, see Fig.~\ref{fig:spec_sn5_1} and Appendix A. The integrated CO images are created by summing up the data cube between the channel $n_i$ and $n_j$ determined by MaxFinder (see Sec. 3.1), maximizing the SNR of the line and the visibility of the detection in integrated images.
We emphasize that all detections are unresolved in the integrated CO images in which the contours are only used to highlight the SNR level and the optical counterpart. The non-Gaussian contours of some sources are due to stacking of pointings with different resolutions which are not indicative for morphologies and do not affect the peak flux measurements on which the following analysis is based.

In Fig.~\ref{fig:spec_sn5_1}, we present spectra of the 17 sources with SNR$>5$ and their optical images overlaid with line intensity contours in which 16 of them have explicit counterparts in the optical/near-infrared imaging. The remaining source, i.e., COALAS-SW.05, is serendipitously found at the edge of pointing SWpoint1 with CO(1-0) SNR=5.9 at $z_{spec}=2.189$. This source is not covered by the VLT nor by HST imaging. However, we find a tentative counterpart in the Subaru $z$-band image \citep{Koyama2013cluster}. Although there is high noise level at the edge of the pointing, it is unlikely to be spurious given the high SNR, thus we include it in the catalogue.

\begin{landscape}
\begin{table}
\vspace{-0.5cm}
\caption{\label{tab:atcasnr5} ATCA CO(1-0) detections}
\vspace{-0.3cm}
\begin{tabular}{cccccccccccc}
\hline\hline
  \multicolumn{1}{c}{ID} &
  \multicolumn{1}{c}{$z_{\rm spec,opt}$}  &     
  \multicolumn{1}{c}{$z_{\rm CO}$} &
  \multicolumn{1}{c}{$I_{\rm CO}$} &
  \multicolumn{1}{c}{SNR$_{peak}$} &
  \multicolumn{1}{c}{FWHM} &
  \multicolumn{1}{c}{$L'_{\rm CO(1-0)}$} &
  \multicolumn{1}{c}{Counterpart} &
  \multicolumn{1}{c}{$f_{\rm flux}$} &
  \multicolumn{1}{c}{$C$} & 
  \multicolumn{1}{c}{$R_{\rm SoFiA}$} &
  \multicolumn{1}{c}{Category} \\
 &    & & [Jy~km~s$^{-1}$] & & [km~s$^{-1}$] & [$10^{10}~{\rm K~km~s^{-1}~pc^2}$] & &\\
\hline
COALAS-SW.01   &   2.1478   & 2.1480 & 0.15$\pm$0.02 & 8.1 & 339$\pm$41 & 3.5$\pm$0.5  & I,Ks,HAE & 1.0 & 1.0 & 1.0 & A\\
COALAS-SW.02   &    2.1612    & 2.1618 & 0.13$\pm$0.02 & 7.5 & 515$\pm$67 & 3.1$\pm$0.5 & I,Ks,HAE & 1.0 & 1.0 & 1.0 & A\\
COALAS-SW.03  &    2.1618   & 2.1627 & 0.16$\pm$0.03 & 6.6 & 353$\pm$52 & 3.8$\pm$0.7 & I,Ks,HAE & 1.1 & 0.9 & 1.0 & A\\
COALAS-SW.04    &   2.1701    & 2.1642 & 0.22$\pm$0.04 & 6.0 & 588$\pm$94 & 5.1$\pm$1.0 & I,HAE & 1.0 & 0.8 & -- & B\\
COALAS-SW.05    &    --   & 2.1890 & 0.51$\pm$0.10 & 5.9 & 371$\pm$62 & 12.5$\pm$2.5 & $z^*$ & 1.0 & 0.9 & 1.0 & A\\
COALAS-SW.06    &   2.1670    & 2.1630 & 0.17$\pm$0.03 & 5.9 & 759$\pm$159 & 3.9$\pm$0.8 & I,Ks,HAE & 1.0 & 0.9 & 1.0 & A\\
COALAS-SW.07    &   --    & 2.1218 & 0.14$\pm$0.03 & 5.6 & 852$\pm$162 & 3.3$\pm$0.7 & I,Ks,HAE & 1.0 & 0.9 & -- & B\\
COALAS-SW.08    &   --    & 2.1033 & 0.28$\pm$0.06 & 5.4 & 213$\pm$30 & 6.5$\pm$1.5 & I,Ks & 1.0 & 1.0 & 0.9 & A\\
COALAS-SW.09 &  2.1510   & 2.1530 & 0.15$\pm$0.03 & 5.4 & 422$\pm$83 & 3.5$\pm$0.8 & I,Ks,HAE & 1.0 & 0.8 & 1.0 & A\\
COALAS-SW.10     &    --   & 2.1533 & 0.13$\pm$0.03 & 5.3 & 615$\pm$120 & 3.1$\pm$0.7 & I,Ks & 1.0 & 0.9 & 0.8 & A\\
COALAS-SW.11   &   --     & 2.1590 & 0.22$\pm$0.05 & 5.3 & 868$\pm$185 & 5.3$\pm$1.2 & I & 0.9 & 0.9 & 1.0& A\\  
COALAS-SW.12     &   --    & 2.1532 & 0.17$\pm$0.04 & 5.3 & 260$\pm$47 & 4.1$\pm$0.9 & I,Ks,HAE & 1.0 & 0.9 & 1.0 & A\\
COALAS-SW.13    &   --    & 2.1892 & 0.16$\pm$0.04 & 5.2 & 656$\pm$44 & 3.9$\pm$0.9 & I,Ks,HAE & 1.0 & 0.8 & 0.9 & A\\ 
COALAS-SW.14    &   --    & 2.1354 & 0.10$\pm$0.02 & 5.2 & 227$\pm$43 & 2.4$\pm$0.5 & Ks & 1.0 & 0.9 & 1.0 & A\\
COALAS-SW.15     &   KMOS    & 2.1567 & 0.06$\pm$0.02 & 5.1 & 268$\pm$49 & 1.4$\pm$0.4 & I,Ks,HAE & 1.6 & 0.1 & 1.0 & A\\ 
COALAS-SW.16    &   --    & 2.2017 & 0.12$\pm$0.03 & 5.0 & 544$\pm$114 & 2.8$\pm$0.7 & I,Ks,HAE & 1.2 & 0.6 & 1.0 & A\\
COALAS-SW.17     &   --    & 2.1519 & 0.50$\pm$0.12 & 5.0 & 413$\pm$83 & 11.7$\pm$2.8 & I & 1.5 & 0.3 & 0.9 & A\\
\hline
COALAS-SW.18   &    --    & 2.1801 & 0.20$\pm$0.05 & 4.9 & 757$\pm$156 & 4.8$\pm$1.2 & I & 1.6 & 0.1 & 1.0 & A\\
COALAS-SW.19    &   --    & 2.1674 & 0.11$\pm$0.03 & 4.9 & 845$\pm$195 & 2.6$\pm$0.6 & I & 1.3 & 0.4 & 0.7 & A\\ 
COALAS-SW.20    &   --    & 2.1282 & 0.27$\pm$0.07 & 4.9 & 648$\pm$142 & 6.4$\pm$1.6 & I,Ks & 1.1 & 0.6 & 1.0& A\\
COALAS-SW.21    &   --    & 2.1237 & 0.30$\pm$0.07 & 4.9 & 1274$\pm$290 & 6.9$\pm$1.7 & Ks & 1.4 & 0.2 & 0.9 & A\\
COALAS-SW.22    &   --    & 2.1702 & 0.21$\pm$0.05 & 4.8 & 762$\pm$168 & 5.0$\pm$1.2 & I & 1.3 & 0.4 & 0.8 & A\\
COALAS-SW.23    &   --    & 2.1629 & 0.17$\pm$0.04 & 4.8 & 471$\pm$101 & 4.0$\pm$1.0 & I & 0.9 & 0.8 & 0.7 & A\\
COALAS-SW.24    &   --    & 2.1395 & 0.16$\pm$0.04 & 4.8 & 313$\pm$71 & 3.6$\pm$0.9 & I & 1.9 & 0.1 & 1.0 & A\\
COALAS-SW.25    &   --    & 2.1680 & 0.16$\pm$0.04 & 4.7 & 623$\pm$87 & 3.8$\pm$1.0 & I & 1.1 & 0.5 & 0.9 & A\\
COALAS-SW.26    &   --    & 2.1897 & 0.22$\pm$0.06 & 4.7 & 334$\pm$70 & 5.4$\pm$1.4 & I,Ks & 1.2 & 0.3 & 1.0 & A\\
COALAS-SW.27    &   --    & 2.1691 & 0.25$\pm$0.06 & 4.7 & 333$\pm$72 & 5.9$\pm$1.5 & I & 1.1 & 0.6 & 1.0 & A\\
COALAS-SW.28    &   --    & 2.0950 & 0.31$\pm$0.08 & 4.6 & 322$\pm$65 & 6.9$\pm$1.8 & I & 1.6 & 0.1 & 0.9 & A\\
COALAS-SW.29    &   --    & 2.2125 & 0.12$\pm$0.03 & 4.6 & 283$\pm$58 & 2.9$\pm$0.8 & I,Ks & 1.7 & 0.1 & 1.0 & A\\
COALAS-SW.30    &   --    & 2.1315 & 0.14$\pm$0.04 & 4.6 & 326$\pm$68 & 3.2$\pm$0.9 & I & 1.1 & 0.8 & 0.9 & A\\
COALAS-SW.31 &   --    & 2.2034 & 0.12$\pm$0.03 & 4.5 & 279$\pm$60 & 3.0$\pm$0.8 & I,Ks & 2.1 & 0.1 & 1.0 & A\\
COALAS-SW.32     &  --    & 2.1750 & 0.09$\pm$0.03 & 4.5 & 219$\pm$47 & 2.3$\pm$0.6 &  I,Ks & 1.6 & 0.1 & 0.8 & A\\ 
COALAS-SW.33     &    --   & 2.1824 & 0.11$\pm$0.03 & 4.4 & 204$\pm$44 & 2.6$\pm$0.7 & I & 1.5 & 0.2 & 0.9 & A\\
COALAS-SW.34     &    --   & 2.1205 & 0.14$\pm$0.04 & 4.4 & 686$\pm$171 & 3.2$\pm$0.9 & I & 1.9 & 0.1 & 0.6 & A\\
COALAS-SW.35    &   --    & 2.1912 & 0.10$\pm$0.03 & 4.4 & 245$\pm$52 & 2.4$\pm$0.6 & I & 1.6 & 0.2 & 1.0 & A\\
COALAS-SW.36 &   --    & 2.1558 & 0.08$\pm$0.02 & 4.4 & 217$\pm$46 & 1.8$\pm$0.5 & I & 1.3 & 0.4 & 1.0 & A\\
COALAS-SW.37 &   --    & 2.1355 & 0.13$\pm$0.03 & 4.4 & 447$\pm$108 & 3.0$\pm$0.8 & Ks & 1.1 & 0.7 & 0.7 & A\\
COALAS-SW.38     &   --    & 2.1462 & 0.21$\pm$0.06 & 4.3 & 1313$\pm$302 & 4.9$\pm$1.4 & Ks & 1.5 & 0.1 & 1.0 & A\\ 
COALAS-SW.39    &   --    & 2.2148 & 0.24$\pm$0.07 & 4.3 & 189$\pm$42 & 6.1$\pm$1.7 & I & 2.2 & 0.1 & 0.6 & A\\
COALAS-SW.40  &   --    & 2.1229 & 0.08$\pm$0.02 & 4.2 & 442$\pm$120 & 1.8$\pm$0.5 & I & 2.4 & 0.1 & 1.0 & A\\
COALAS-SW.41    &   --    & 2.1617 & 0.18$\pm$0.05 & 4.2 & 290$\pm$65 & 4.2$\pm$1.2 & I,HAE & 1.7 & 0.2 & 0.6 & A\\
COALAS-SW.42    &   --    & 2.1431 & 0.10$\pm$0.03 & 4.1 & 219$\pm$47 & 2.4$\pm$0.7 & I & 2.1 & 0.1 & 0.9 & A\\
COALAS-SW.43   &   --    & 2.1284 & 0.13$\pm$0.04 & 4.0 & 229$\pm$57 & 3.0$\pm$0.9 & I,Ks & 1.5 & 0.2 & 1.0 & A\\
COALAS-SW.44    &   --    & 2.1906 & 0.13$\pm$0.04 & 4.0 & 557$\pm$158 & 3.2$\pm$1.0 & I,Ks & 2.1 & 0.1 & 0.9 & A\\
COALAS-SW.45    &   2.1606   & 2.1603 & 0.06$\pm$0.02 & 4.4 & 168$\pm$37 & 1.4$\pm$0.4 & I,Ks,HAE & 0.3 & 1.2 & -- & C\\ 
COALAS-SW.46    &   KMOS   & 2.1621 & 0.10$\pm$0.03 & 3.8 & 567$\pm$154 & 2.4$\pm$0.8 & I,Ks,HAE & -- & -- & -- & C\\
\hline
\end{tabular}\\
{Note: {$z_{\rm spec,opt}$: optical redshifts from \cite{Shimakawa2018SW}, KMOS: VLT/KMOS redshift from P\'{e}rez-Mart\'{i}nez et al. (in prep.)}; Counterpart: I (HST F814W), Ks (VLT), $z$ (Subaru), HAE (\citealt{Koyama2013cluster,Shimakawa2018SW}); 
$f_{\rm flux}$ and $C$ are flux factor and completeness, respectively; $R_{\rm SoFiA}$: $Reliability$ from SoFiA. $*$ Tentative counterpart. {The horizontal line highlights sources with with SNR$>$5.}
}
\end{table}
\end{landscape}

\subsection{Line identification and redshift distribution}
\label{sub:lineidentification}
In this study, all detected lines are identified as CO(1-0) at $z\sim2.2$. As shown in Table~\ref{tab:atcasnr5}, nine sources are found with consistent redshifts to $z_{\rm CO(1-0)}$ from near-infrared spectroscopy presented in \cite{Shimakawa2018SW} and/or the VLT/KMOS observations in P\'erez-Mart\'inez et al (in prep., private consultation). Regarding the remaining sources without a secure spectroscopic redshift, in case of higher-J CO transitions, e.g., CO(2-1) at $z=5.3$, we inspected Herschel/SPIRE images for all CO detections and found no red colors that could suggest redshifts beyond $z=4-5$ \citep[e.g.,][]{Riechers2017}. On the other hand, adopting the number density of $z>5$ submillimeter galaxies in COLDz survey \citep{Riechers2020density}, the expected number of $z>5$ galaxies is less than 1 in this field. Thus, the detection of CO(2-1) appears very unlikely and thus cannot be used to constrain CO(2-1) space density at $z=5.3$. We compute the CO(1-0) luminosities $L'_{\rm CO(1-0)}$, following \cite{Solomon1997}, namely,
\begin{equation}
    L'_{\rm CO(1-0)} = 3.25\times 10^7 S\Delta{v}\frac{D_{L}^{2}}{(1+z)^3\nu_{obs}^{2}}   ~{\rm K~km~s^{-1}~pc^2}
\end{equation}
where $S\Delta{v}$ is the integrated flux of the line in Jy~km~s$^{-1}$ (corrected for flux boosting, see Section~\ref{subsec:fluxboosting}), $D_L$ is the luminosity distance in Mpc, and $\nu_{obs}$ is the observed frequency.

\begin{figure}
\centering
\includegraphics[width=0.49\textwidth]{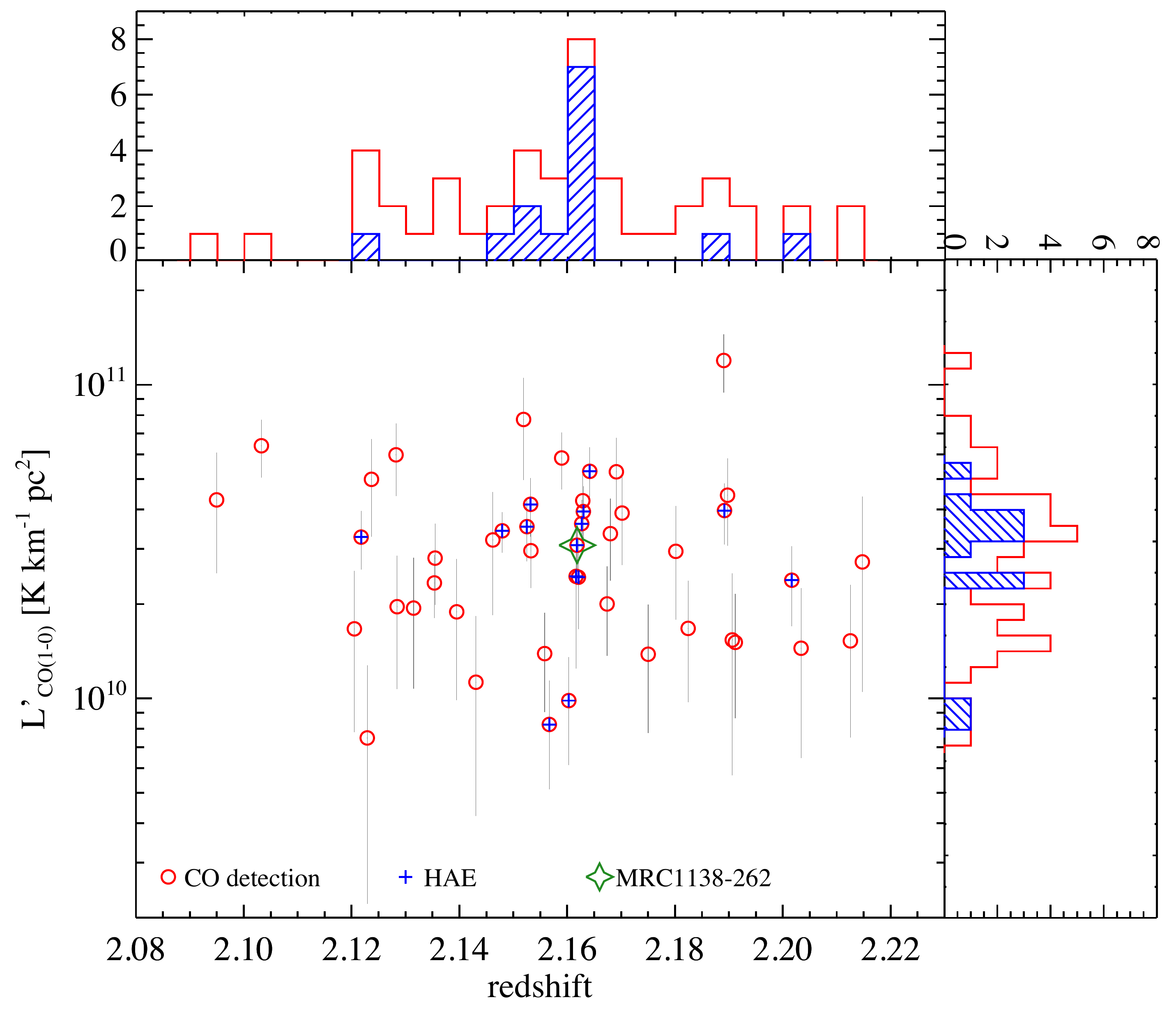}
\caption{
CO luminosities vs. redshifts for CO detections in this study. Top and right panels show the redshift and CO luminosity distribution of our detections, respectively. In histograms, all CO emitters are shown in red, and HAEs with CO detections are shown in blue. 
\label{fig:hist}
}
\end{figure}

In Fig.~\ref{fig:hist}, we show the redshift and CO(1-0) luminosity distribution, respectively. The detections spread over the redshift range $z=$2.09--2.22 with CO(1-0) luminosity up to $L'_{\rm CO(1-0)}=2\times10^{11}$ ${\rm K~km~s^{-1}~pc^2}$. Previous membership was based on H$\alpha$ narrow-band imaging and subsequent near-infrared spectroscopy. Thus, the HAE redshift range of $z$=2.146--2.170 is a selection effect of the narrow band filter $NB_{2071}$ \citep{Shimakawa2018SW} which is limited within the redshift range of $z=2.15\pm0.2$ and unable to probe the wider structure of Spiderweb cluster. 
Strikingly, the CO emitters show an overdensity from $z=2.12$--2.21, a factor of 3.8 wider in velocity than the range traced by the HAEs. 
Such an overdensity remains at $z=2.12$--2.21 even if we limit the $z_{\rm CO}$ histogram to only those sources with SNR$>5$. It is independent of the rms level, line width and other properties, which are robust and unlikely to be impacted by selection effects or miss-identifications. 
Therefore, this indicates that this structure has a scale of 120 co-moving Mpc (cMpc), suggesting a filament and/or a super structure. We discuss this further in  Sec.~\ref{subsec:supercluster}.

 \subsection{Spatial distribution of CO emitters}
 \label{sub:spatical}
 
\begin{figure}
\centering
\includegraphics[width=0.49\textwidth]{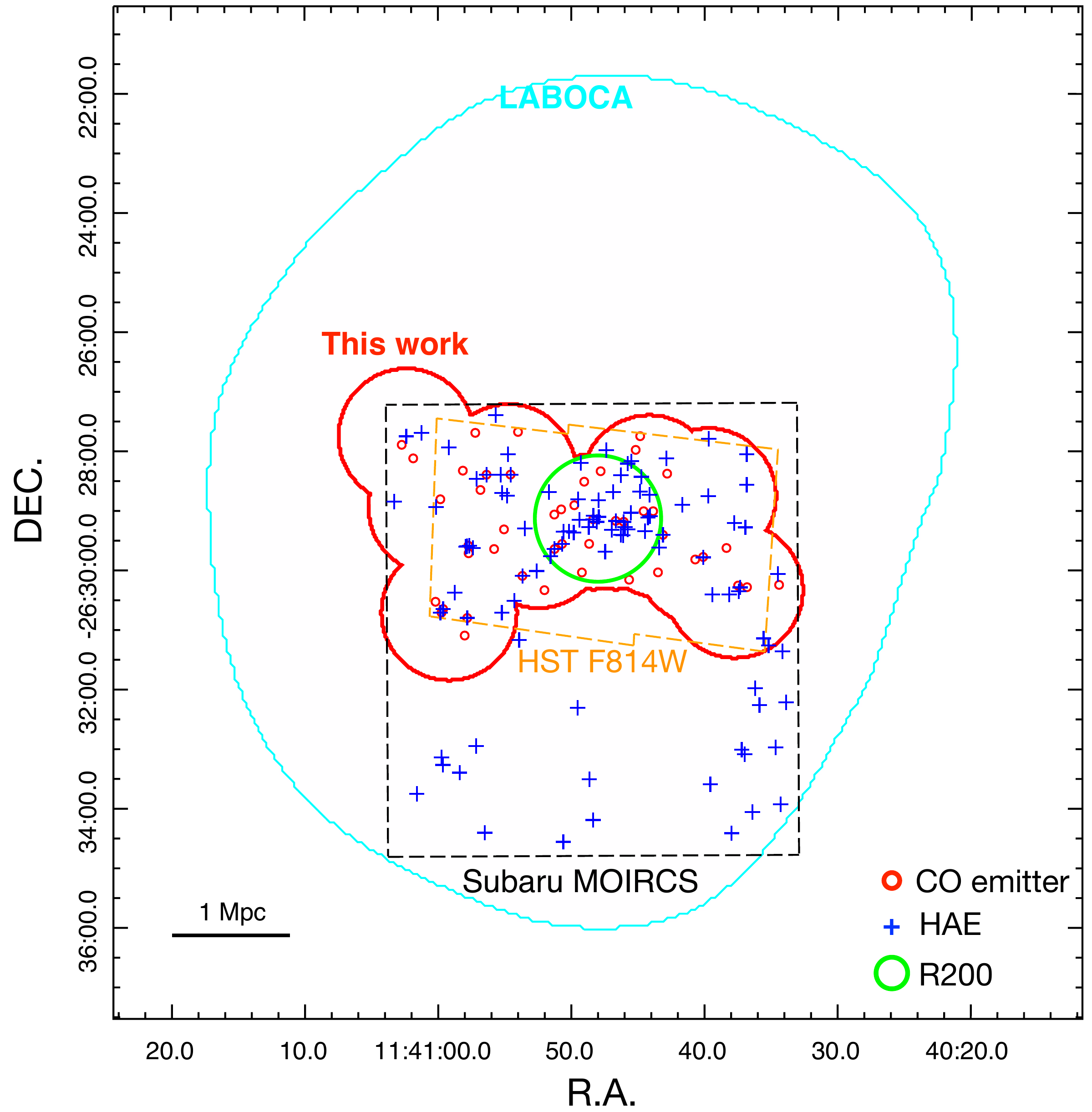}
\caption{
The footprints of this work and previous studies, overlaid by CO sources (red circles) in this work and HAEs (blue crosses) in \cite{Koyama2013cluster} and \cite{Shimakawa2018SW}.  The $R_{200}$ radius \citep{Shimakawa2014HAE} is shown in a green circle centering on the MRC1138-262.
\label{fig:footprint}
}
\end{figure}

In Fig.~\ref{fig:footprint}, we show the footprints of previous studies, including HST I band image, Subaru MOIRCS Ks band image \citep{Koyama2013cluster,Shimakawa2014HAE} and LABOCA 870$\mu$m continuum map. In this work, using ATCA we observed an area slightly larger than the HST F814W image, covering the majority of LABOCA sources in \cite{Dannerbauer2014LABOCA} and HAEs in \cite{Koyama2013cluster} and \citep{Shimakawa2014HAE}. As the $R_{200}$ shown in green circle, the central cluster core around MRC1138 is well covered by all observations, while the large filamentary structure traced by HAEs is still not fully observed with ATCA.
 
\begin{figure*}
\centering
\includegraphics[width=0.99\textwidth]{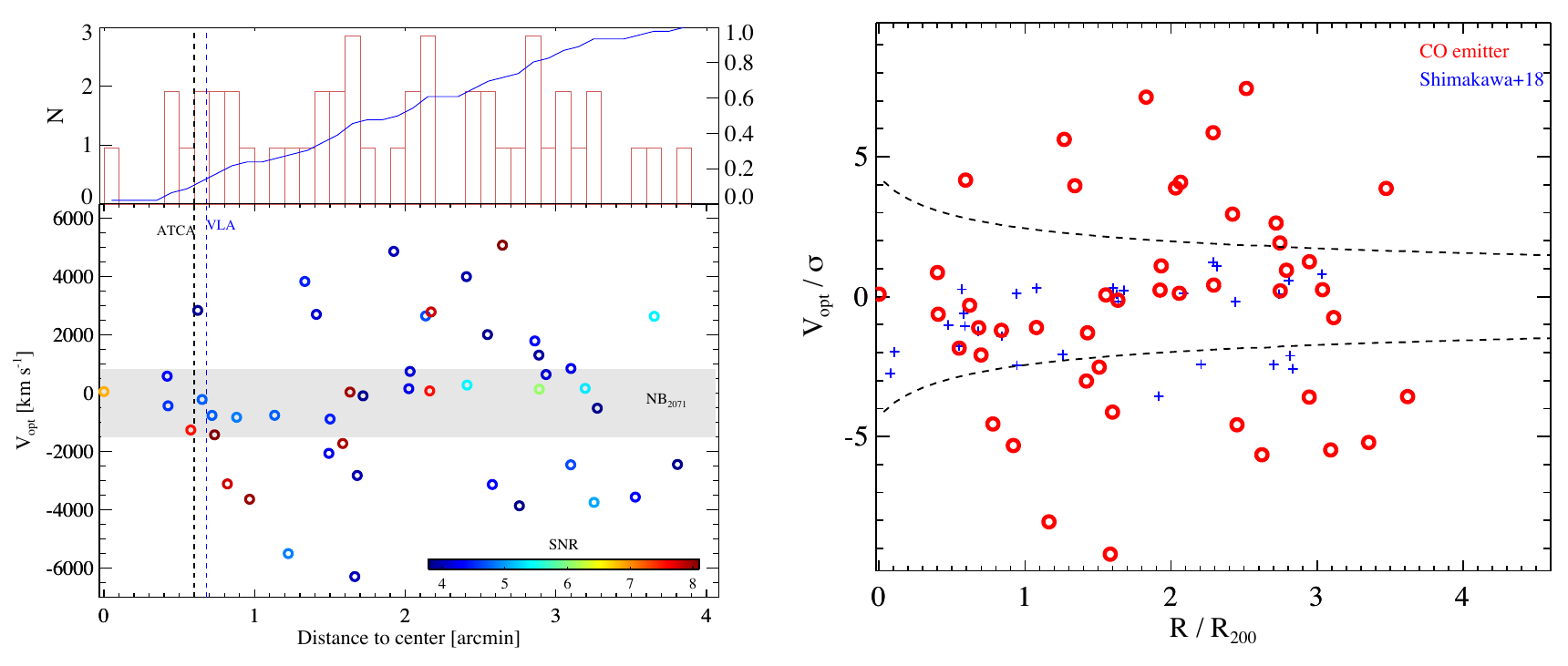}
\caption{
{\it Left panel:} Distribution of CO emitters vs. their sky distances to the central radio galaxy MRC1138-262. Top panel shows histogram (red) and cumulative fraction (blue) of CO emitters as a function of the distances to the center galaxy. In the bottom panel, CO emitters are color-coded with their SNR. Grey shaded area marks the redshift range observed by the narrow band $N_{2071}$ \citep{Shimakawa2018SW} of MOIRCS at Subaru. Blue and black dashed lines mark the field of view of VLA and ATCA single pointing, respectively.
{\it Right panel:} Normalized velocity vs. distance to center galaxy for CO emitters in this study and HAEs with a spectroscopic redshift \citep{Shimakawa2018SW}. The normalization factors are $\sigma=683$~km~s$^{-1}$ and $R_{200}=0.53$~Mpc (Table 2 in \citealt{Shimakawa2014HAE}). The two dashed curves show the escape velocity (positive and negative) as a function of distance from the cluster mass $1.71\times10^{14}~M_{\odot}$ \citep{Shimakawa2014HAE} using prescription in \cite{Rhee2017}. 
Half of the CO emitters in this work have velocities outside of the region enclosed by these two curves which are not expected to be gravitationally bound to the central mass.
\label{fig:velo_dist}
}
\end{figure*}

In Fig.~\ref{fig:velo_dist}, we show the sky and velocity distributions of CO emitters in this study. Respectively, the left panel shows the velocity of CO emitters to their sky distance to the central radio galaxy MRC1138-262 while the right panel presents the normalized version.
In Fig.~\ref{fig:velo_dist}-left, the CO emitters scatter largely in view of the observer, i.e., line-of-sight velocity range of $\pm6500$ km~$s^{-1}$ and sky distances of 0$'$.3--4$'$.0 to the central galaxy. 
Robust sources (SNR~$>5$) are also found at large velocity (e.g., 5500~km/s) and large distance (up to 3$'$.7) to the center, which further strengthens the large structure indicated by the large redshift range $z=2.12-2.21$ of the CO overdensity (Sec. \ref{sub:lineidentification}).
Meanwhile as indicated by the histogram in Fig.~\ref{fig:velo_dist}-left, 90\% of the CO emitters are found to have large distances $0'.5-4'$ to the central radio galaxy. Thus the single pointings of VLA or ATCA observations at 7mm focusing on the center radio galaxy only covers $10\%$ of CO sources presented in this work. 
Regarding the field of view of ALMA, only the ALMA compact array (ACA) at 3~mm can have comparable primary beam (FWHM $\sim1'.7$) as the VLA and ATCA ones. Thus it is also impractical to discover large structures similar as the Spiderweb protocluster in single pointing mode.
Therefore, surveys with large area ($>20$ arcmin$^2$) are essential to discover the majority of CO emitters in similar structures as found in this study. 
On the other hand, narrow-band imaging can cover large area \citep{Koyama2013cluster,Shimakawa2018SW} but the redshift range probed covers only $18\%$ of the velocity range observed by ATCA which is thus unable to discover such large structure.

Given the large velocity and spacial offsets from the center galaxy, some CO sources do not appear to be gravitationally bound to the center galaxy. 
In Fig.~\ref{fig:velo_dist}-right, we present the normalized velocity and distances to the center for CO emitters in this work and HAEs with spectroscopic redshifts (\citealt{Shimakawa2018SW}). 
\cite{Shimakawa2014HAE} calculated a mass of $1.71\times10^{14}~M_{\odot}$ for the Spiderweb protocluster assuming virialization in the core. Using this mass, galaxies with velocities that fall within the region enveloped by the two curves \citep{Rhee2017} are expected to be gravitationally bound. 
We find that 21 CO emitters have velocities inside the region enclosed by the two curves. In contrast, the remaining CO sources scatter largely in the diagram with velocities faster than the escape velocities which are thus unlikely to be bound with the cluster core.
They concluded that the two groups are not gravitationally bound to the core system and will remain distributed within the overall galaxy cluster. 
In this work, the velocities of CO emitters out of the bound region are 1--5 times larger than the escape velocities. 
Meanwhile, these unbound CO emitters are mostly at large distance (88\% have $R>R_{200}$), only four sources (including the center radio galaxy) are found within the $R_{200}$ radius. This is in line with the picture drawn in \cite{Shimakawa2014HAE} where the center is already collapsed and nearly virialized. The outer regions appears still highly structured and some CO emitters could be at the early phase of assembly towards the protocluster core.

\subsection{Comparison to previous surveys}
This catalogue contains 46 robust CO(1-0) detections which is the largest catalogue of CO emitters in a galaxy cluster field, and among the largest ones published for one contiguous field ever. In Table~\ref{tab:comparison}, we compared this catalogue to CO surveys from the literature in both high-z galaxy (proto)clusters and blank fields. To keep consistency, we show only the number of detections that are found with counterparts in high resolution images (e.g., HST, ALMA and VLA).
In terms of the observed area, the 21 arcmin$^2$ covered by ATCA is the largest one ever observed in protocluster fields to date while the observed area of other surveys in protocluster fields are smaller by factors of 3--10. The number of 46 detected sources in Spiderweb field is larger than the numbers presented in other protocluster work by a  factor of 2--15. Thus the CO emitter sample in this study constitutes the largest CO sample in a protocluster to date. We note that CO emitters in other protoclusters could be underestimated. As mentioned in Sec.~\ref{sub:spatical}, it is possible that other high-z protoclusters also contain a large number of CO emitters but the majority of them would be missed due to the small area observed.
This sample size is also comparable with respect to surveys in blank fields, e.g., COLDz \citep{Pavesi2018COLDz,Riechers2020COLF}, ASPECS \citep{Gonzalez-Lopez2019ASPECS,Decarli2019COLF} and PHiBSS2 \citep{Lenkic2020COLF}. 
These CO surveys in blank fields have been used to constrain the CO luminosity function. Similarly, the large number of CO emitters with a significant overdensity in the Spiderweb cluster enables us to constrain the CO luminosity function and molecular gas density in an overdense region for the first time.

\begin{table*}
{
\centering
\caption{CO surveys in high-z clusters and blank fields}
\label{tab:comparison}
\centering
\begin{tabular}{cccccccc}
\hline
\hline
  \multicolumn{1}{c}{Project} &
  \multicolumn{1}{c}{Facility} &
  \multicolumn{1}{c}{Field} &
  \multicolumn{1}{c}{$z$} &
  \multicolumn{1}{c}{CO transition} &
  \multicolumn{1}{c}{$N_{\rm det,ctp}$} &
  \multicolumn{1}{c}{Area} &
  \multicolumn{1}{c}{Reference$^{a}$}\\
  & & & & & & [arcmin$^2$] & \\
\hline
(High-z clusters) \\
 COALAS & ATCA  & Spiderweb  & 2.16  & J=1 & 46  & 21.3 &  This study \\
 -- & VLA  & CL J1001$^b$  & 2.51  &  J=1 & 14  & 1.3 &  W16,18 \\
 -- & ALMA  & XMMXCS J2215.9-1738$^{c}$ & 1.46  &  J=2 & 17  & 2.3 &  H17 \\
 -- & ALMA  & SpARCS-J0225 & 1.60  &  J=2 & 13  & 2.6 & N17,19 \\
 -- & ALMA, VLA  & Cl J1449+0856 & 1.99  &  J=1,3,4 & 8  & 1.5 & C18 \\
 -- & ALMA  & DRC & 4.00  &  J=6 & 10  & $1.0$ & O18 \\
 -- & ALMA  & SPT2349-56 & 4.30  &  J=4 & 8 & 0.2 &  M18 \\
 -- & ALMA  & Spiderweb & 2.16  &  J=3 & 4  & 0.6 &  T19 \\
 -- & ALMA  & USS1558-003 & 2.53  &  J=3 & 6 & 0.6 &  T19 \\
 -- & ALMA  & 4C23.56 & 2.49  &  J=3 &  6  & 0.6  &  T19 \\
 -- & ALMA, VLA  & HXMM20 & 2.60  &  J=1,3 & 5 & 0.8 &  GG19 \\
 -- & ALMA  & HELAISS02 & 2.17  &  J=3 & 3  &0.6  &  GG19 \\
 -- & ALMA  & 4C23.56 & 2.49  &  J=4 &  11  & 0.5 &  L19 \\
 -- & ALMA  & SPT2349-56 & 4.30  & J=4, [CII] & 24  & 7.2 &  H20 \\
 \hline
 (Blank fields) \\
  ASPECS-pilot & ALMA  & HUDF & --  &  J=2--4,7,8 & 9 & 1.2  &  D16 \\
  COLDz & VLA  & GOODS-N,COSMOS & --  &  J=1,2 & 17 & 59.8 &  P18, R19 \\
 ASPECS & ALMA  & HUDF & --  &  J=2--4 & 18 & 4.6 &  GL19, D19 \\
 PHIBSS2 & PdBI  & $d$ & --  &  J=2--6 & 43 & 130.0 &  L20 \\
\hline\hline\end{tabular}\\}
\tablefoot{
$N_{\rm det,ctp}$: Number of detection with counterpart.

$^a$ W16,18: \cite{Wang_T2016cluster,WangTao2018CO}; H17: \cite{Hayashi2017CO21_cluster}; N17,19: \cite{Noble2017cluster,Noble2019cluster}; C18: \cite{Coogan2018}; O18: \cite{Oteo2018cluster}; M18: \cite{Miller2018cluster_z4}; T19: \cite{Tadaki2019cluster}; GG19: \cite{Gomez-Guijarro2019}; L19: \cite{Lee2019cluster}; GL19: \cite{Gonzalez-Lopez2019ASPECS}, D16: \cite{Decarli2016COLF}; D19: \cite{Decarli2019COLF}; L20: \cite{Lenkic2020COLF}, H20: \cite{Hill2020cluster}.

$^b$ \cite{Gomez-Guijarro2019} identified six of the 14 emitters as tentative detections. Champagne et al. (submitted) indicate a different explanation for the CO emitters using new VLA observations.

$^c$ This cluster is also studied by \cite{Stach2017cluster} who found 6 CO(2-1) emitters with shallower ALMA observations.

$^d$ COSMOS, GOODS-N and EGS.}

\end{table*}

\subsection{CO(1-0) line width versus luminosity}
\label{subsec:tully}
The relationship between CO luminosity, and CO line width is an analogue of the Tully-Fisher relation \citep{TF1977} which can be interpreted as a relationship between the molecular gas mass and the velocity necessary for centrifugal support of a rotating disk \citep{Bothwell2013}. Although showing a large scatter, this relation is a practical tool that enables us to indicatively diagnose the size and mass of gas reservoirs without high resolution imaging.

\begin{figure}
\centering
\includegraphics[width=0.49\textwidth]{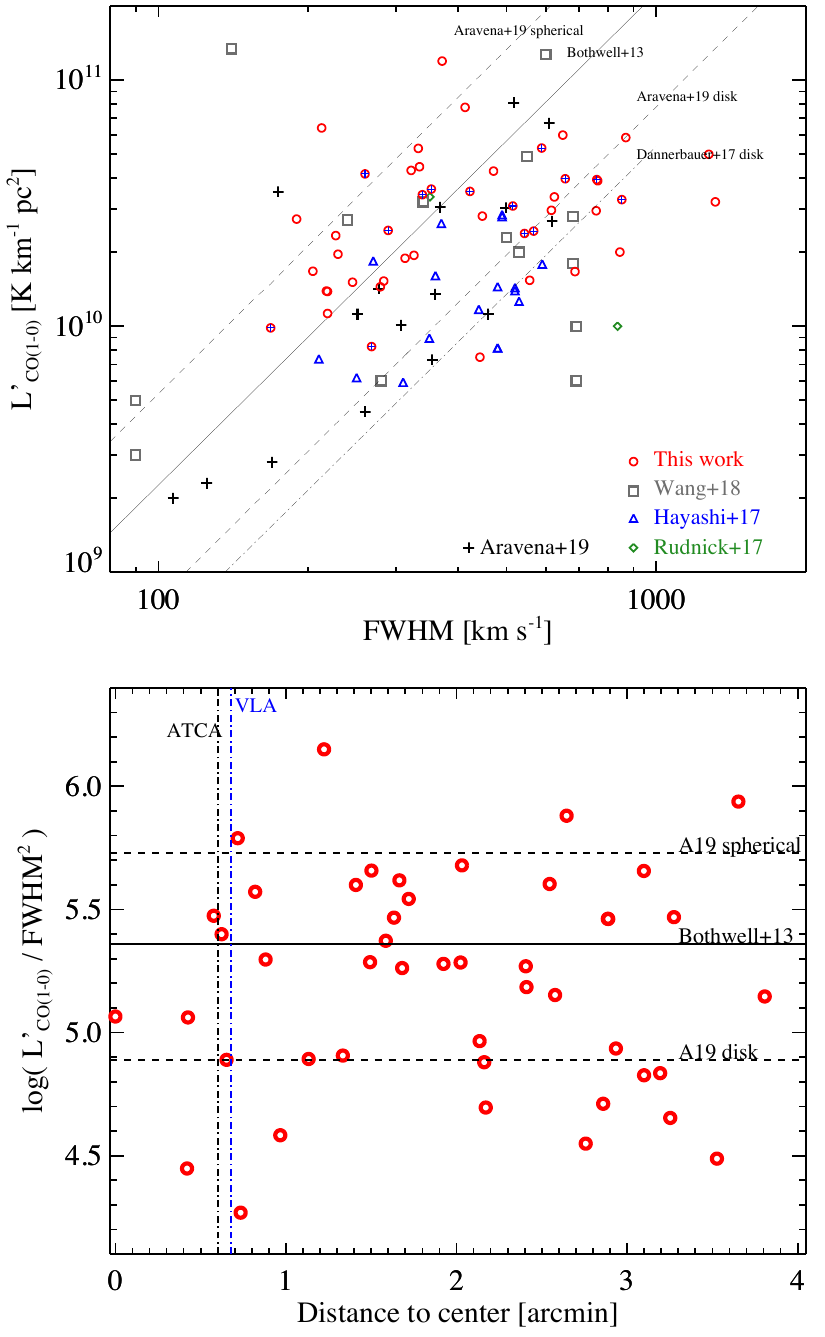}
\caption{
{\it Top panel:} CO luminosity vs. FWHM. CO detections in this study are shown in red circles with HAEs marked by blue crosses. The dashed lines are from \cite{Aravena2019ASPECS} where the upper line represents a spherical model and the bottom line a disk model. ASPECS galaxies \citep{Aravena2019ASPECS} are shown in black pluses which lie mostly along the disk model line. The solid line shows the relation from \cite{Bothwell2013}, dot dashed line presents the disk model from \cite{Dannerbauer2017disk}.
{\it Bottom panel:} CO luminosity normalized by FWHM$^2$ vs. distance to the center galaxy. Vertical lines mark the half FWHM of ATCA (black) and VLA (blue) primary beams, respectively.
This figure shows that CO emitters in Spiderweb cluster are more likely to be starbursts (mergers) compared to other high-z clusters, and these starbursts are beyond the FoV of the single pointing focusing on the central galaxy.
 \label{fig:TF}
}
\end{figure}

In Fig.~\ref{fig:TF}-top, we show the CO(1-0) luminosity $L'_{\rm CO}$ versus the full-width-half maximum (FWHM) of our detections. In addition, we include low-J CO observations, tracing the cold molecular gas, of high-z galaxy (proto)clusters: CO(2-1) observations of a cluster at $z=1.46$ \citep{Hayashi2017CO21_cluster}, CO(1-0) observations of two clusters at $z=1.62$ and 2.51 \citep{Rudnick2017cluster,WangTao2018CO}, respectively. The CO(2-1) luminosities are converted into CO(1-0) luminosities using the brightness temperature ratio $r_{21}$=0.76 \citep{Daddi2015}. 
CO emitters in \cite{WangTao2018CO} and \cite{Hayashi2017CO21_cluster} cover a wide parameter space on the plot, while other samples are limited to smaller luminosity and FWHM range due to small sample size and shallow depths.
We show the parametrization of the $L'_{\rm CO(1-0)}$ vs. FWHM relationship for few representative cases: the relation for SMGs between these two parameters established by \cite{Bothwell2013} (radius = 7~kpc and $\alpha_{\rm CO}=1$), a disk model with a radius of 13~kpc from \cite{Dannerbauer2017disk}, a disk galaxy and a spherical model described in \cite{Aravena2019ASPECS}. The four lines have the same slope but different normalizations due to the assumption of different disk radius and $\alpha_{\rm CO}$. The relation of spherical models (i.e., with small radius and low $\alpha_{\rm CO}$) is above other models in larger disk size and high $\alpha_{\rm CO}$. Compact galaxies with low $\alpha_{\rm CO}$ (e.g., starbursts) tend to have higher CO luminosity at fixed FWHM while the $L'_{\rm CO(1-0)}$--FWHM relation of typical disk galaxies tend to be lower than the the relation of compact starbursts.

Interestingly, we find that a dozen of CO sources show higher ratios of CO luminosity to line width compared to other galaxy cluster samples which are above the solid line for disk model (Fig.~\ref{fig:TF}-top) from \cite{Bothwell2013}. Some members of the Spiderweb protocluster lie close and/or even above the `spherical' model line presented in \citet{Aravena2019ASPECS}. Compared to other high-z cluster members, our cluster members have a larger CO luminosity at specific line width than the samples from \cite{Hayashi2017CO21_cluster}, \cite{Rudnick2017cluster} and \cite{WangTao2018CO}. 
Overall, we find larger CO luminosities at fixed line width than all other previous surveys. We note that the optical/near-infrared images (see Fig~\ref{fig:spec_sn5_1} and Appendix A) show that the morphologies of our cluster members are complex and almost half of them appear irregular. The fit to the spherical CO gas models and their irregular morphology together suggest that these galaxies are merger-like systems, consistent with the expected high merger rate in galaxy cluster environments at high redshift \citep{Coogan2018}. Therefore, their $L'_{\rm CO}$ appears similar to SMGs suggesting that this cluster is rich of SMGs (and mergers) with respect to other high-z clusters. This further strengthens the conclusion presented in \cite{Dannerbauer2014LABOCA}.

In Fig.~\ref{fig:TF}-bottom, we normalized the CO luminosity by FWHM$^2$ as an indicator of starburstiness, and show their sky distance to the center galaxy. Intriguingly, the most starburst-like galaxies, i.e., sources above the spherical model in \cite{Aravena2019ASPECS}, locate at $>0'.5-3'.7$  away from the center galaxy. These galaxies would be totally missed in single pointing observations at 7mm with ATCA and VLA, if the phase center is focusing on the center cluster galaxy. Given that other surveys in protocluster fields are all observing a small area, the majority of starbursting members in protoclusters would be missed if the starbursting members are also far away from the center as in the Spiderweb. In this case, the starburst/merger rates in high-z protoclusters could be significantly underestimated due to the limited field of view. This work indicates that conventional single pointing observations could have severe bias in protocluster fields that missed the bulk of starbursting members.

\subsection{CO(1-0) luminosity function}

\label{subsec:luminosityfunction}
Our large CO(1-0) sample enables us to establish the CO luminosity function and study the molecular gas density in a galaxy protocluster environment for the first time. We construct the CO luminosity function adopting Eq. 2 from \cite{Decarli2016COLF}:
\begin{equation}
    \Phi({\rm log}{L_i}) = \frac{1}{V_{c}}\sum_{j=1}^{N_i}\frac{R_j}{C_j}
\end{equation}
where $N_i$ is the number of galaxies that fall within the luminosity bin  log$L_i$ -0.25 and log$L_i$ + 0.25, $V_c$ is defined as the observed co-moving volume, $C_j$ is the completeness described in Section~\ref{subsec:fluxboosting} and $R_j$ is the reliability. Note that \cite{Lenkic2020COLF} introduced two association probabilities to weight each detection which are the probability of an optical counterpart associated with a particular CO transition $P_{z,j}$ and the probability of a line detection spatially associated with an optical counterpart $P_{a,j}$. However, in this study, $P_{z,j}$ equals to 1 as all lines are identified as CO(1-0), and $P_{a,j}$ is close to 1 because we used tolerance $<3''$ to match any optical counterpart and the uncertainty of spatial association is negligible with respect to 5--13$''$ beam sizes. Thus we stick to the equation by \cite{Decarli2019COLF}. 
We adopt $R=1$ for a source with a counterpart and a matching of a previously known/obtained spectroscopic redshift. In cases of sources without a spectroscopic redshift, we conservatively treat the reliability from SoFiA as upper limits. In each luminosity bin, the CO space densities are created 1000 times, each time the reliability $R_j$ is distributed to a random value that is uniformly distributed between 0 and such upper limit (see also \citealt{Gonzalez-Lopez2019ASPECS,Pavesi2018COLDz,Riechers2019COLDz}). Line candidates are kept in our analysis only if the random value is below the reliability threshold. The final CO luminosity functions are the medians of all the CO LF realizations and the uncertainty is computed as Poissonian uncertainty $\sqrt{N_i}$. 

\begin{figure*}
\centering
\includegraphics[width=0.98\textwidth]{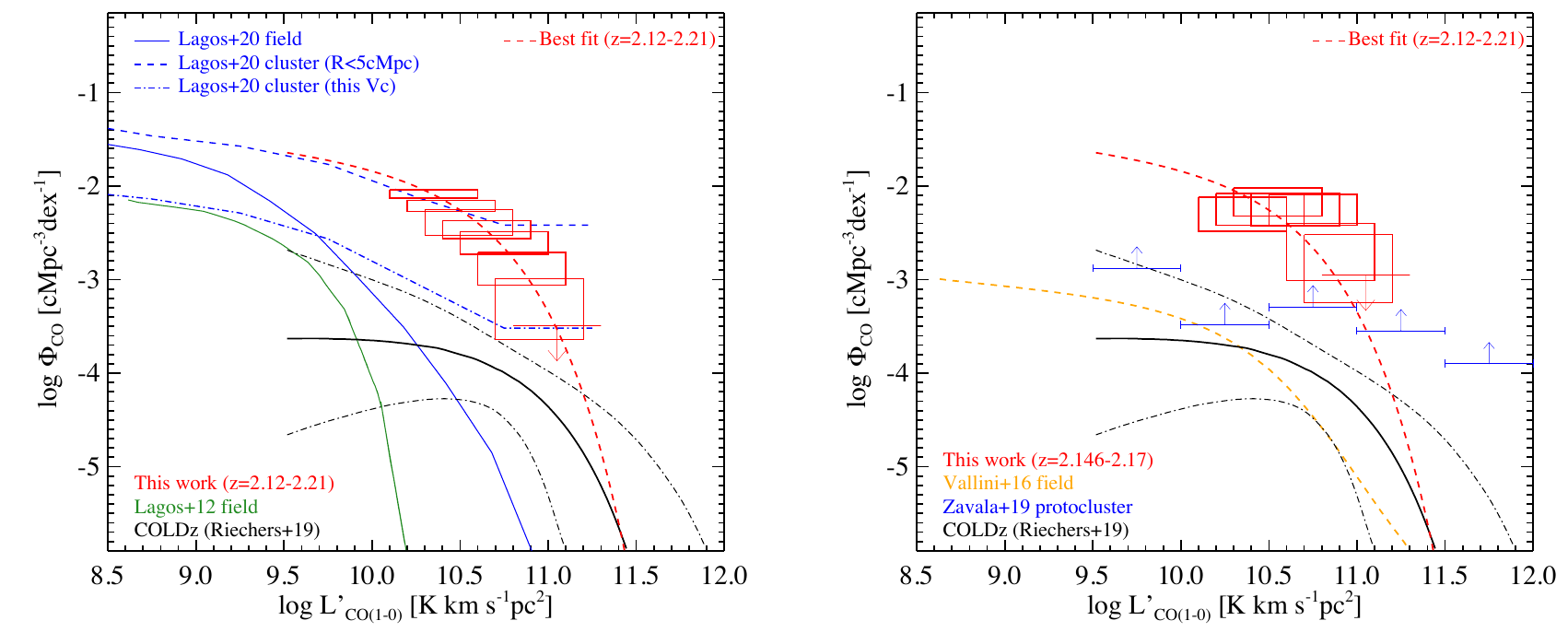}
\caption{
The CO(1-0) luminosity functions for the Spiderweb cluster and samples from the literature. {\it Left:} CO(1-0) luminosity density in each luminosity bins at $z=2.12$--2.21 in this work, compared with the results at $z\sim2$ from the COLDz \citep{Riechers2019COLDz}, and predictions from semi-analytic SHARK models \citep{Lagos2012COLF,Lagos2020}. The dotted-dashed black lines mark the 5th and 95th percentile of the COLDz LF. Red dashed curve shows the best fit to our data. We show CO luminosity functions for protoclusters derived from SHARK models within two different volumes. One (blue dashed curve) is calculated in a spherical volume with radius $R<5$~cMpc and the other one (blue dot-dashed curve) is in the identical volume to where the observed CO LF is measured, respectively.
{\it Right:} Same as the left panel, but boxes are shown for CO luminosity density in a narrow redshift range $z=2.146$--2.17. Blue arrows mark the lower limits for protoclusters in \cite{Zavala2019cluster}, and orange curve shows the results from \cite{Vallini2016COLF}.
\label{fig:COLF}
}
\end{figure*}

In order to calculate the volume, first we define the redshift range of this cluster as $z=2.146-2.17$ (i.e., 33 co-moving Mpc or $\pm$1200 km/s) being determined by the redshift range of the HAEs in this cluster (Fig.\ref{fig:hist}). This redshift range is also consistent with the $z=2.15-2.17$ used in \cite{Dannerbauer2014LABOCA} for calculating the SFRD, which enable us to consistently compare the CO space density to the SFRD. 
The total sky coverage of the ATCA mosaic is 25.8 arcmin$^2$. We exclude shallow areas that have rms noise $>$1 mJy per beam where no source is detected there, thus we adopt 21.3 arcmin$^2$, i.e., 53 cMpc$^2$. We also compute the CO luminosity function for detections at the redshift range $z=2.12-2.21$ (124 cMpc) as an overdensity of CO emitters is also seen within this redshift range (Fig.~\ref{fig:hist}). 
Therefore, the observed co-moving volume is $\sim6600$ cMpc$^3$ for the redshift range $z=2.12-2.21$, and $\sim1650$ cMpc$^3$ for $z=2.146-2.17$, respectively.
We show the CO number densities for the two redshift ranges in Fig.~\ref{fig:COLF} and Table~\ref{t5}. 
The CO luminosity function is calculated in luminosity bins of 0.5 dex width to compare consistently with the COLDz LF, starting log$(L'_{\rm CO}/{\rm K~km~s^{-1}~pc^2})=10.1$ in steps of 0.1 dex. The log$(L'_{\rm CO}/{\rm K~km~s^{-1}~pc^2})<10.1$ bins are ignored because of the small sample size and high incompleteness.
The CO space density at $z=2.146-2.17$ is consistent with the ones at $z=2.12-2.21$ within error bars (Fig.~\ref{fig:COLF}-right).
Note that the narrow redshift range $z=2.146-2.17$ is also comparable with the gravitationally bound region in Fig.~\ref{fig:velo_dist}-right, the similar amplitudes of CO LFs in the two redsfhit ranges indicate that the CO space density is not enhanced in the gravitationally bound region of the cluster core.
Given that the larger number of sources found in the wide redshift range can better constrain the CO luminosity function than done with the narrower redshift range, we thus adopt the CO luminosity function at $z=2.12-2.21$ for further analysis. 

Furthermore, we verified the CO luminosity function for robust sources (SNR$>$5) at $z=2.12-2.21$. We found that the median amplitude of CO LF would be scaled down by 0.5 dex due to having less galaxies per luminosity bin but still one order of magnitude higher than that for COLDz and agrees with the density of the full sample within error bars. On the other hand, given that SNR$>$5 sources are robustly detected, using their $Reliability$ as upper limits appears too conservative and would overkill, we thus directly adopt their $Reliability$ from SoFiA as the $R_j$ in Eq. 5, and found that the resulted CO LF for $>5\sigma$ sources is in excellent agreement with the best fit (Fig.~\ref{fig:COLF}-left). Therefore, These confirms that our results of CO luminosity function are robust and even conservative.

\begin{table}
\caption{\label{t5} CO(1-0) luminosity function}
\renewcommand\arraystretch{1.4}
\begin{tabular}{c|c|c|c|c}
\hline\hline
     \multicolumn{1}{c|}{} & \multicolumn{2}{c|}{$z=2.146$--2.17}  & \multicolumn{2}{c}{$z=2.12$--2.21}  \\ 
\hline
  \multicolumn{1}{c|}{log$L'_{\rm CO}$} &
  \multicolumn{1}{c|}{ $N$  } &
  \multicolumn{1}{c|}{log$\Phi_{\rm CO}$} &
  \multicolumn{1}{c|}{  $N$  } &
  \multicolumn{1}{c}{log$\Phi_{\rm CO}$} \\
\hline
10.1--10.6 & 12 & [-2.12, -2.48] & 28 & [-2.04, -2.13]\\
10.2--10.7 & 13 & [-2.08, -2.42] & 27 & [-2.15, -2.27] \\
10.3--10.8 & 16 & [-2.02, -2.32] & 26 & [-2.25, -2.53]\\
10.4--10.9 & 14 & [-2.08, -2.43] & 22 & [-2.37, -2.56]\\
10.5--11.0 & 12 & [-2.09, -2.42] & 18 & [-2.49, -2.73]\\
10.6--11.1 & 6 & [-2.40, -3.01] & 10 & [-2.71, -2.06]\\
10.7--11.2 & 4 & [-2.52, -3.25] & 6 & [-2.99, -2.64]\\
10.8--11.3 & 1 & $<$ -2.95  & 2 & $<$-3.49\\
\hline\hline
\end{tabular}
\end{table}

We fit the observed CO(1-0) luminosity function (LF) of $z=2.12-2.21$ with a Schechter function \citep{Schechter1976LF}, in the logarithmic form used in \cite{Riechers2019COLDz},
\begin{equation}
    {\rm log}\Phi = {\rm log}\Phi_{*}+\alpha~{\rm log}\frac{L'}{L'_{*}}-\frac{1}{{\rm ln}~10}\frac{L'}{L'_{*}}+{\rm log}({\rm ln}~10)
\end{equation}
where $\Phi_{*}$ is the scale number of galaxies per unit co-moving volume, in units of
cMpc$^{-3}$~dex$^{-1}$, $L'_{*}$ is the scale line luminosity in units of K~km~s$^{-1}$~pc$^{2}$ (the ``knee'' of the luminosity function) and $\alpha$ is the slope of the faint end. We show the best fitting results for CO LF at $z=2.12$--2.21 in Table~\ref{tab:fitting}, including one fit with free parameters and two with fixed slopes. The best fit with free Schechter parameters has $\alpha=-0.21\pm1.13$, log$(L'_{*}/{\rm K~km~s^{-1}~pc^2})=$10.48$\pm$0.38 and log$\Phi_{*}=-2.16\pm0.49$. We note that the slope of the faint end of the LF, $\alpha$, is very uncertain due to the limited number of galaxies. This parameter is sensitive to the corrections we applied for reliability and completeness. 

In order to consistently compare our result with studies from the literature, we fit the CO LF with fixed values of $\alpha$ in our analysis, adopting $\alpha=0.08$ from the COLDz CO(1-0) LF \citep{Riechers2019COLDz} and $\alpha=-0.2$ from the ASPECS project \citep{Decarli2019COLF}, respectively. As listed in Table~\ref{t6}, the best fit with fixed $\alpha=0.08$ shows log$(L'_{*}/{\rm K~km~s^{-1}~pc^2})=$10.39$\pm$0.06 and log$\Phi_{*}$=-2.06$\pm$0.07. The log$L_{*}$ is close to the COLDz log$(L'_{*}/{\rm K~km~s^{-1}~pc^2})=$10.7 (50th percentile) while the CO number density $\Phi_{*}$ in this study is 2.6$\pm$0.5 dex higher than COLDz number density. If we adopt $\alpha=-0.2$ as in \cite{Decarli2019COLF}, we would get log$(L'_{*}/{\rm K~km~s^{-1}~pc^2})=$10.48$\pm$0.07 and log$\Phi_{*}=-2.16\pm0.09$, -- the density is still more than one order of magnitude higher than the estimated one in COLDz. Comparing with number densities $\Phi_{*}$ from ASPECS \citep{Decarli2019COLF}, our results are higher than ASPECS CO(3-2) density at $z\sim2.6$ by 1.4$\pm$0.5 dex. 
To summarize, the fits with fixed slopes show a consistent scaled line luminosity log$(L'_{*}/{\rm K~km~s^{-1}~pc^2})\sim10.5$ and number density (log$\Phi_{*}\sim-2.2$) to the fit with free parameters. All fits on our sample show more than 1.5 dex higher number density of CO emitters than the COLDz and ASPECS surveys.

\begin{table}
\caption{\label{t6} Schechter function fit parameter constraints to CO(1-0) luminosity function at $2.12<z<2.21$}
\renewcommand\arraystretch{1.4}
\begin{tabular}{cccc}
\hline
  \multicolumn{1}{c}{$\alpha$} &
  \multicolumn{1}{c}{ log$L'_{*}$}  &
  \multicolumn{1}{c}{log$\Phi_{*}$} \\
\hline
-0.21$\pm$1.13 & 10.48$\pm$0.38  & -2.16$\pm$0.49 \\ 
0.08$^*$ &  10.39$\pm$0.06 & -2.06$\pm$0.07  \\  
-0.20$^*$ & 10.48$\pm$0.07  & -2.16$\pm$0.09 \\  
\hline
\end{tabular}\\
{$^*$ The value is fixed in the fitting procedure.}
\label{tab:fitting}
\end{table}

For comparison, we add data from the literature: the observed CO LFs from the field galaxy sample COLDz \citep{Riechers2019COLDz} and \cite{Vallini2016COLF}.
The CO luminosity density in the Spiderweb cluster is remarkably higher than that in field galaxy samples, e.g., 1.3--1.6 orders of magnitude higher than the median value in COLDz survey at log$(L'_{\rm CO}/{\rm K~km~s^{-1}~pc^2})\sim 10.5$, and 3.4 orders of magnitude higher than the results from \cite{Vallini2016COLF}.
Recently, \cite{Zavala2019cluster} provided a constraint on gas content in protoclusters at $z\sim2.0$--2.5 based on ALMA continuum observations which show enhanced gas density in protocluster environment. Our results agree well with the lower limits in \cite{Zavala2019cluster}, only the extrapolation of the Schechter function at log$(L'_{\rm CO}/{\rm K~km~s^{-1}})>$11.
Furthermore, we compare with results from the semi-analytic models in \cite{Lagos2012COLF} and \cite{Lagos2020} predicted for both the random field and the overdense environment. We refer to a detailed discussion on the comparison in Sec.~\ref{sec:discussion_simu}.

\subsection{Molecular gas density in the Spiderweb protocluster}
\label{subsec:moleculardensity}

In general, the cold molecular gas density can be obtained via integrating the CO(1-0) luminosity functions and applying a fixed CO to H$_{2}$ conversion factor $\alpha_{\rm CO}$ for all CO emitters. 
However, the $\alpha_{\rm CO}$ of these CO emitters are uncertain. Some CO emitters appear to be starbursts (see Sec~\ref{subsec:tully}) which have lower $\alpha_{\rm CO}$  than that applied for field galaxies (e.g., $\alpha_{\rm CO}=3.6~M_{\odot}$~(K~km~s$^{-1}$~pc$^{2}$)$^{-1}$  in \citealt{Riechers2019COLDz} and \citealt{Decarli2019COLF}).
In order to have a reasonable conversion to cold molecular gas mass, we apply two typical $\alpha_{\rm CO}$ values for starbursts and normally star-forming galaxies that are diagnosed by the $L'_{\rm CO}$--FWHM diagram, respectively. For the 19 CO sources above the \cite{Bothwell2013} line in Fig~\ref{fig:TF}, we adopt a starburst-like $\alpha_{\rm CO}=0.8~M_{\odot}$~(K~km~s$^{-1}$~pc$^{2}$)$^{-1}$ (\citealt{Emonts2018}), and adopt $\alpha_{\rm CO}=3.6 ~M_{\odot}$~(K~km~s$^{-1}$~pc$^{2}$)$^{-1}$ \citep{Daddi2010SFL} for the rest of sources that appear to be disk-like. Given that \cite{WangTao2018CO} adopted $\alpha_{\rm CO}\gtrapprox 4.0 ~M_{\odot}$~(K~km~s$^{-1}$~pc$^{2}$)$^{-1}$ in the $z=2.5$ starbursting cluster \citep[which are determined following][]{Genzel2015}, the values applied in this work are  conservative. 

In order to calculate the cold molecular gas density, we followed three different methods: (1) We simply sum up the CO(1-0) luminosity of all detections at $z=2.12$--2.21 in our catalogue (taking into account the flux boosting) and convert it into the cold molecular gas using $\alpha_{\rm CO}$ described above. By dividing the total cold molecular gas mass by the co-moving volume, we obtain a cold molecular gas density $\rho{\rm (H_{2})}=5.9\times10^{8}~M_{\odot}$~Mpc$^{-3}$; (2) We derive the cold molecular gas density by adopting a similar method presented in \cite{Decarli2016COLF}, summing up and weighting gas mass of each CO detection by completeness and reliability same as we did for the calculation of the CO luminosity function (Section~\ref{subsec:luminosityfunction}), i.e.,
\begin{equation}
    \rho ({\rm H_2}) = \frac{1}{V}\sum_{j=1}^{N_i}\frac{M_{j}~R_j}{C_j}
\end{equation}where $M_{j}$ is the gas mass the $j$th galaxy calculated as $M_{j}=\alpha_{\rm CO,j}L'_{\rm CO(1-0),j}$.
We obtain a high value for the cold molecular gas density $\rho{\rm (H_{2})}=9.0_{-3.4}^{+3.6}\times10^{8}~M_{\odot}$~Mpc$^{-3}$ with uncertainties of 5th and 95th percentiles; 
(3) Assuming the same CO LF slope as the COLDz work, we derive the cold gas density by scaling up the COLDz CO(1-0) LF to our work. The COLDz project \citep{Riechers2019COLDz} reported a cold molecular gas density of $1.1-5.6\times10^7~M_{\odot}$~Mpc$^{-3}$ at $z\sim2$ assuming $\alpha_{\rm CO}=3.6~M_{\odot}$~(K~km~s$^{-1}$~pc$^{2}$)$^{-1}$. The COLDz CO(1-0) LF (50th percentile) shows log$(L'_{*}/{\rm K~km~s^{-1}~pc^2})$=10.7, log$\Phi_{*}=-3.87$ and $\alpha=-0.08$. Comparing to our Schechter function fit with the same slope (Table~\ref{tab:fitting}), our CO LF is 1.7$\pm$0.06 dex higher in scaled number density and 0.2$\pm$0.05 dex lower in scaled luminosity. The integrated Schechter function is $L_{int}= \Phi_{*}L'_{*}\Gamma(\alpha+2)$. Thus, the cold molecular gas density in this protocluster would be 1.5 magnitude higher than that in the COLDz survey, i.e., $\rho{\rm (H_{2})}= 1.06_{-0.24}^{+0.31}\times10^{9}~M_{\odot}$~Mpc$^{-3}$ adopting the same $\alpha_{\rm CO}=3.6$. This calculation will be scaled down by a factor of 1.3 using the specified $\alpha_{\rm CO}$ described above. We note that the calculations (1) and (2) are based on only the observed part of the luminosity function, undetected galaxies in lower luminosity bins are not accounted for the given large uncertainties in the individual luminosity bins and the unknown intrinsic shape of the CO luminosity function. The three calculations agree well with each other while the estimate (2) appears conservative but consistent with the results of (1) and (3). 
We thus adopt the calculation (2) and estimate the H$_2$ density $\rho{\rm (H_{2})}=$0.56--1.26$\times10^{9}$ M$_\odot$ Mpc$^{-3}$  in the Spiderweb galaxy protocluster.

\begin{figure}
\centering
\includegraphics[width=0.48\textwidth]{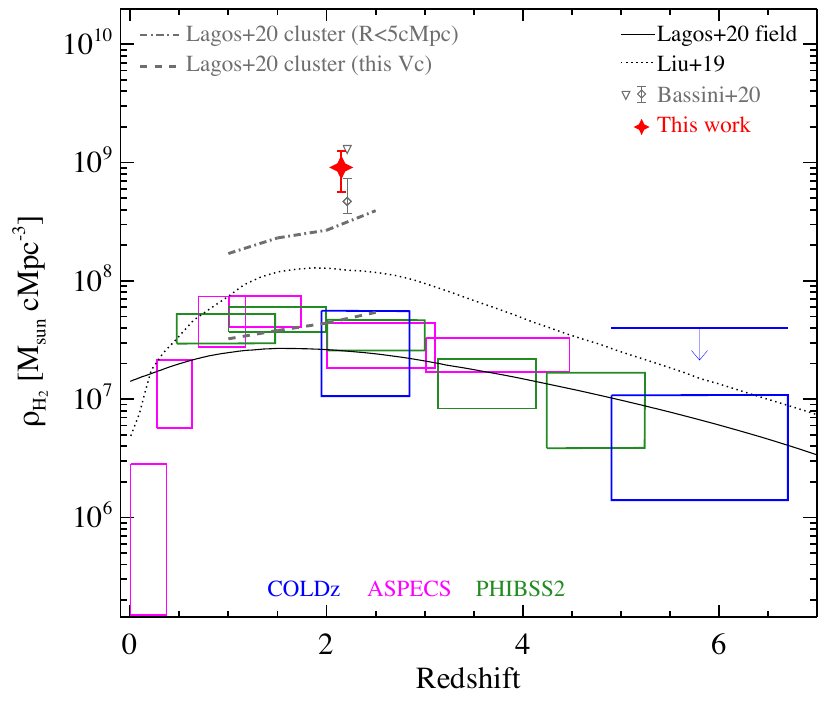}
\caption{
H$_2$ density of this study (the red star) compared to results from COLDz \citep{Riechers2019COLDz}, ASPECS \citep{Decarli2020COLF} and PHIBSS2 \citep{Lenkic2020COLF} overlaid by fitting of field galaxy samples from \cite{Liu2019GasEvolution} and \cite{Lagos2020}. 
Gas density in galaxy clusters from simulations are shown in gray curves for SHARK \citep{Lagos2020} in two volumes, and gray diamond with a bar for DIANOGA \citep{Bassini2020simu}, respectively. The gray diamond with error bar shows the median, and the downward triangle marks the maximum values for cluster structures in DIANOGA simulations. The gas density in this work is significantly higher than observed in blank fields and also higher than predictions for clusters in simulations.}
 \label{fig:molgas}

\end{figure}

In Fig.~\ref{fig:molgas}, we compare the cold molecular gas density with results from the COLDz \citep{Riechers2019COLDz} and ASPECS \citep{Decarli2016COLF,Decarli2019COLF} surveys. Similar to the CO luminosity function, the H$_2$ density in the Spiderweb cluster is two orders of magnitude higher than that determined in field galaxy samples like COLDz and ASPECS. We note that the ASPECS H$_2$ densities \citep{Decarli2019COLF} based on CO(3-2) with assumed $r_{31}=0.42\pm0.07$ should be scaled down by a factor of two on average as calibrated by VLA CO(1-0) followups (\citealt{Riechers2020COLF})
which has been recently updated in \cite{Decarli2020COLF} adopting $r_{31}=0.84\pm0.26$ from \cite{Riechers2020COLF}. The H$_2$ density in this galaxy protocluster is completely based on CO(1-0), the best tracer of the cold molecular gas and free from CO excitation issues.

An important caveat here is that the H$_2$ density strongly depends on the adopted $\alpha_{\rm CO}$. Here, we adopted a bimodal $\alpha_{\rm CO}$. Although this is a good representation of  the full sample, the value of $\alpha_{\rm CO}$ for individual galaxies can be highly uncertain.

\section{Discussion}
\label{sec:discussion}

\subsection{Is the Spiderweb a super-protocluster?}
\label{subsec:supercluster}

The redshift range of the CO emitters ($z=2.12-2.21$) is  a factor of four wider than the redshift range of HAEs and LAEs in this field and indicates a size of 124~Mpc in co-moving distance along the line of sight. 
\cite{Kurk2000} conducted narrow band Ly$\alpha$  imaging in this field using the VLT/FORS. They used a narrow band filter with a central
wavelength of 381.4~nm and a FWHM of 6.5~nm, covering a redshift range of $z=2.111-2.164$.
\cite{Pentericci2000} followed up the LAEs selected from \cite{Kurk2000} with the VLT/FORS in MOS mode. They confirmed reshifts of Ly$\alpha$ emitters mostly at $z=2.14-2.17$ except one quasar at $z=2.183$ with a Ly$\alpha$ absorption feature.
\cite{Kuiper2011} discovered a sample of satellite galaxies around the center galaxy MRC1138-262 via detecting H$_\alpha$, [O II] and [O III] emissions with SINFONI spectroscopy. Their redshifts span $z=2.13$--2.17, larger than the redshift range of narrow-band-selected HAEs \citep{Koyama2013cluster,Shimakawa2018SW}. 
Both the LAEs and HAEs selected in this field are limited on a narrow redshift range due to use of narrow band filters.
The velocity distribution of the confirmed satellite galaxies shows a broad, double-peaked velocity structure with  $\sigma=1013\pm87$~km~s$^{-1}$ which is proposed to be a result of the merger of two subclusters \citep{Kuiper2011}. Including all spectroscopically confirmed HAEs and LAEs, \cite{Shimakawa2014HAE} measured a velocity dispersion of 683~km~s$^{-1}$ in the cluster core with a $R_{200}=0.53$ Mpc in physical distance.
However, in this study, the CO emitters show a much larger spread in redshifts and sky which can not be interpreted as an effect of pure velocity dispersion of cluster galaxies as it would lead to a unrealistically large dispersion exceeding 4000 km~s$^{-1}$. 
Furthermore, since robust sources (SNR$>5$, Fig.~\ref{fig:velo_dist}) are found to be at large velocity (e.g., 5500~km~s$^{-1}$) and large distance (e.g., 3$'$.7), the large spread of velocities and distances to the center cannot be explained as a selection effect.
Thus we argue that the large redshift range of CO emitters is tracing a large scale structure that is significantly wider than the assumed previously one. 
Overall, the CO emitters provide us a very different view of the protocluster structure than that of Ly$\alpha$ and H$_\alpha$ emitters, and appear to be better tracers of the large scale matter density. 

The $\sim$120 Mpc co-moving size is significantly larger than the scale of any known galaxy (proto)clusters at low and high redshift, beside the supercluster Hyperion \citep{Cucciati2018} that is found with a radial size of 150~cMpc. 
Note that the overdensity of CO emitters at $z=$2.12--2.21 is only slightly higher than the density of the remaining redshift range covered by the ATCA. This structure could be larger than what we observed and more comparable to the radial scale of the Hyperion. This might indicate that we observe a galaxy super-protocluster similar as the Hyperion \citep{Cucciati2018}. 
However, given that the 53~cMpc$^2$ area covered by ATCA is 72 times smaller than the 60$\times$60~cMpc$^2$ size of the Hyperion, possibly this structure could be a protocluster connected to a large scale filamentary structure along line of sight. We are not able to distinguish if this structure is a filament or a super-protocluster with the current data in hand. In case of a super-protocluster, such large structure appears unlikely to totally collapse by $z=0$, given that half of the CO emitters in this study appears not bound with the central mass (Fig.~\ref{fig:velo_dist}-right). This also holds for the Hyperion. However, filaments can act as highways to fuel the inner regions of clusters with gas and galaxies, as indicated by the ThreeHundred simulations \citep{Rost2020simu}. There, filaments around clusters can extend out to five $R_{200}$ that are comparable with the distances between the CO emitters and the center core in this study. Thus the collapse of the gas and dark matter would be accelerated if it is a protocluster connected to a large scale filament.

{The volume observed in this work is very elongated, $\sim6\times9\times120$~cMpc$^3$, thus it is difficult, with such a narrow field of view, to decide whether the structure may be filamentary.}
In order to confirm the filamentary or super-protocluster structure, a comprehensive redshift survey covering large area is highly desired. 
\cite{Emonts2018} report 6 CO(4-3) and 4 [CI](1-0) detections using ALMA observations around the MRC1138$-$262. These observations took only 106 minutes integration time (in a single ALMA pointing). Very recently,\cite{Valentino2020ALMACO} show the impressive capabilities of ALMA to obtain CO and [CI] transitions in a few minutes of on-source integration time for a sample of a few tens high-z IR-selected galaxies. 
In the Spiderweb protocluster field, high resolution ALMA observations of alternative molecular gas tracer, e.g., [CI] (accompanying CO(4-3) line emission), would be invaluable to constrain SFEs of individual galaxies and the gas excitation via constructing CO SLEDs. 
The detection rate in Spiderweb field would be more complete with ALMA given the high brightness of CO(4-3) with respect to CO(1-0). The bandwith of ALMA band 4 is twice wider than the 2~GHz bandwidth of ATCA.  However, given that the CO(4-3) line is at a $4\times$ higher frequency than CO(1-0), the redshift range probed by ALMA band 4 is just half of the range probed by ATCA, thus being less efficient to probe the large scale of this structure. 
Therefore, benefiting from its large field of view and broad bandwidth, ATCA is still the best facility for CO line surveys with the largest volume. In the near future, the upgraded Graphics Processor Units\footnote{https://theconversation.com/a-brain-transplant-for-one-of-australias-top-telescopes-129138} on ATCA will double the amount of bandwidth that can be observed, thus future ATCA observations of the Spiderweb field will be crucial to unveil the edge of this large structure.

\subsection{Comparison to simulations}
\label{sec:discussion_simu}

In Fig.~\ref{fig:COLF}, we compared our results with CO LFs in the SHARK semi-analytic models for both field and cluster environment \citep{Lagos2020}. For the field, we compare with GALFORM \citep{Lagos2012COLF} and SHARK \citep{Lagos2018Shark,Lagos2020}. To compare with our protocluster, we search for halos with masses $>10^{13.5}~M_\odot$ in the latest SHARK simulations \cite{Lagos2020} and select the CO(1-0) emitters that fall within the same volume (and geometry) of the Spiderweb protocluster. We find that the CO LF of this protocluster is characterized by a higher number density than the SHARK predictions for both the field and clusters.
Compared to the COLDz CO LF, SHARK \citep{Lagos2020} predicts a CO LF with a different shape, which has a lower number density at bright luminosities, $>10^{10.5}$~K~km~s$^{-1}$~pc$^2$ and higher number density at low luminosities. 
The CO LF of Spiderweb has a more similar shape that resembles the SHARK one for the field but with a remarkably higher number density by 2.3 dex at log$(L'_{*}/{\rm K~km~s^{-1}~pc^2})$=10.5.
To make a fairer comparison with the Spiderweb, we investigated the CO LF of high mass halos in SHARK using two different volumes: (1) a spherical volume of $R<5$~cMpc, which is representative of cluster cores, marked by the blue dashed curve; and (2) the identical volume and geometry of the observed co-moving volume (6600~cMpc$^3$) in this work, marked by blue dot-dashed curve. As shown in Fig.~\ref{fig:COLF}-left, the CO LF of SHARK in the small volume ($R<5$~cMpc core) is comparable to the observed one in the Spiderweb protocluster but  appears to have a slightly too high number density at the brightest end (though with poor statistics it is hard to assess whether this is a true tension with the prediction).
However, when we study the CO LF adopting the same comoving volume as we use for the Spiderweb, the number density decreases by 0.6 dex across all luminosities. 
This indicates that cold molecular gas reservoirs in the simulated SHARK clusters are overdense in center cluster cores while much lower in the outer regions. This could indicate that this protocluster is in a higher density region of the Universe than the average halo of mass $>10^{13.5}~M_{\odot}$ at $z=2$ in SHARK.

In Fig.~\ref{fig:molgas}, we compared our cold molecular gas densities with predictions for clusters from the SHARK semi-analytical models \citep{Lagos2020} and the DIANOGA hydrodynamical simulations \citep{Bassini2020simu}, respectively. 
The core region ($R<5$~cMpc) of massive halos in SHARK shows a median gas density of $2.51_{-0.75}^{+0.76}\times10^8~M_{\odot}~{\rm cMpc^{-3}}$ at $z=2$ with a median $\alpha_{\rm CO}$ of 1.23, indicating a high density of starburst-like members in the center core. This gas density in core region is still lower than the bottom limit of our result by a factor of 2.
In the identical volume and geometry to the one analyzed in this paper (this $V_c$),
the SHARK models predict a molecular gas density of $\rho{\rm (H_2)}=4.40_{-0.75}^{+0.76}\times10^7~M_{\odot}~{\rm cMpc^{-3}}$ at $z=2$ , which is lower than the observed one by 1.3 dex even accounted for uncertainties of $\alpha_{\rm CO}$.
In the DIANOGA simulations, which consist of a set of 12 cosmological simulations of massive galaxy clusters (see \citealt{Bassini2020simu} for further details), there is no direct information on the molecular gas content. We thus compare to the cold gas content that directly fuel the star formation of cluster galaxies.
Between the 12 simulations, we selected the one with the highest gas density at $z\sim2$. This is a simulation of a massive cluster with a mass $M_{200}> 10^{15}~M_{\odot}$ at $z=0$. Therefore, this simulation is suitable for this comparison, since numerical simulations suggest that the Spiderweb complex is the progenitor of a very massive galaxy cluster at $z=0$ (e.g., \citealt{Saro2009cluster}). For this simulated cluster, we chose the snapshot nearest to the redshift of the Spiderweb protocluster, and computed the gas density considering all the cold gas bound to the galaxies within the same volume and geometry.  
By employing $\sim10000$ randomly chosen lines of sight, we find a median density of $\rho=4.7^{+2.6}_{-1.0}\times10^8~M_{\odot}~{\rm cMpc^{-3}}$ (16th and 84th percentiles) with a maximum of $1.29\times10^9~M_{\odot}~{\rm cMpc^{-3}}$ for massive structures with $M_{500}>10^{15}~M_{\odot}$ at $z=2.16$ \citep{Saro2009cluster}. The median gas density agrees with the observed one within error bar and the maximum density is very comparable with the upper limit of our results.

Most of the simulations and semi-analytical models underpredict the star formation rate in protoclusters at $z\sim$2 and 4 (e.g., \citealt{Granato2015simu,Bassini2020simu,Lim2020simu}). This work shows that the molecular gas density in protoclusters can be also underestimated in the SHARK simulations. Regarding the tension between the observations and the SHARK models, one possibility would be that the Spiderweb protocluster is a much higher density
than the one typically traced by halos of masses $>10^{13.5}M_\odot$ in simulations and hence we are not comparing like with like. 
In the future we will investigate if any of the massive halos in these simulations can produce the correct density of star-forming galaxies (and which cosmological volume is required to see such high densities). After isolating whether those systems exist in the simulations, follow-up investigations will focus on comparing star formation efficiencies and their spatial distribution within the protocluster environment. On the other side, the DIANOGA simulations contain more massive structures similar to the Spiderweb protocluster providing very comparable gas density predictions with respect to our observations. We note that even though the total gas densities are comparable, the gas mass function could be different between the observations and simulations. A detailed study on gas mass function requires a deeper inspection which is beyond the scope of this study.
To summarize, this comparison shows that observed high-z super-protocluster and large filamentary structures do exist in cosmological simulations and the consistency with observations provide  encouraging prospects for future simulation studies.

\subsection{SFRD vs. gas density}
Based on submm imaging with APEX-LABOCA, \cite{Dannerbauer2014LABOCA} report that the SFRD of this structure is $1500~M_{\odot}$~yr$^{-1}$~Mpc$^{-3}$ in physical volume, corresponding to a SFRD$=47~M_{\odot}$~yr$^{-1}$~cMpc$^{-3}$ in co-moving volume after applying a scaling factor of $(1+z)^3$. This SFRD is higher than the CSFRD in random fields (see a review in \citealt{Madau2014a}) by 2.6 dex.

Intriguingly, our observations suggest that the H$_2$ density is $1.6\pm0.5$ orders of magnitude higher than in the field which is lower than the 2.6 orders of magnitude excess of SFRD with respect to that in blank fields.
The offset could be due to a combination of several reasons: 
1) a high SFE of the cluster members as suggested by the high fraction of dusty starbursts in this structure; 
2) overestimate of the SFRs due to blending as the dusty SEDs in \citet{Dannerbauer2014LABOCA} are based on low-resolution Herschel and LABOCA (sub)mm photometry that could boost the SFR measurements; 3) the H$_2$ density could be underestimated due to the lack of constraint at faint end of CO LF; and
4) we might be missing some CO emitters in noisy regions. 
In order to better constrain the $L_{\rm IR}$ and SFRs of individual galaxies, deblended photometry (e.g., \citealt{Jin2018cosmos,Liu_DZ2017}) and wide and deep spectral-line imaging (by ATCA and ALMA) would be essential to identify the main contributors to the difference between the SFRD and the H$_{2}$ density of this structure.

\subsection{The importance of large area surveys with wide velocity range}

As listed in Table~\ref{tab:comparison}, previous cold molecular gas surveys often observed a small area around the protocluster (core). 
In contrast, the Spiderweb protocluster exhibits large distances between the most CO emitters and the center galaxy (Fig.~\ref{fig:velo_dist}), indicating that such large structures can be only revealed by large area line surveys.
Other high-z protoclusters could have similarly large spread of gas reservoirs as the Spiderweb protocluster, however, most of CO emitters in such structures would be missed by surveys that covers small area and narrow velocity range, resulting in an underestimate of the total gas mass and gas density. 
Clearly, large area surveys ($>$10 armin$^2$) are indispensable to unveil complete cluster members and unbiased gas content in high-z protoclusters. The area needs to be even larger for super structures like the Spiderweb and the Hyperion.
However, blind line survey appears to the be inefficient and risky, as the existence of CO emitters is unknown in outskirts of protocluster. 
To improve the efficiency, the single dish (sub)mm observations are valuable, e.g., APEX/LABOCA, JCMT/SCUBA2, IRAM 30m/NIKA2 and the upcoming LMT/TolTEC. All these instruments can trace the dust content in protoclusters with large field of view.
As demonstrated in this work and \cite{Hill2020cluster}, the ATCA and ALMA followup of LABOCA submillimeter sources around the cluster core will increase the efficiency of line survey.
On the other hand, the CO(1-0) detection rate of HAEs in this field is $17\%$, which is comparable with that reported by \cite{Tadaki2014cluster}.
Thus CO line followup observations covering most of the rest-frame UV/optical line emitters appear also a good strategy for future surveys. 
A detailed analysis on CO(1-0)-detected HAEs in this field will be presented in future work (Jin et al. in prep.).

\section{Conclusion}
\label{sec:conclusion}
We present new ATCA CO observations of the $z=2.16$ Spiderweb galaxy protocluster field. We find 46 robust CO(1-0) emitting sources at the redshift of the Spiderweb protocluster. For the first time we place constraints on the CO luminosity function and the cold molecular gas density on a galaxy protocluster environment in the distant Universe. Our findings can be summarized as follows:

\spb 475 hours integration time was spent on observations of the CO(1-0) transition of members of the Spiderweb protocluster at $z=2.16$.
We produce a large mosaic of 13 pointings at 7mm, covering an area of 25 armin$^2$ and velocity range of $\pm7000$ km/s. Using multiple source extraction methods, we reveal 46 solid CO(1-0) detections with SNR$>4$ and counterparts in HST optical and/or VLT near-infrared images. 

\spb The CO emitters span a redshift range of $z=2.09-2.22$. We find a CO overdensity at $z=2.12-2.21$ which is four times larger than the velocity range traced by HAEs in previous studies and suggesting a large scale filament or a galaxy super-protocluster in this field. 

\spb We find that 90\% of the CO emitters are $>0'.5-4'$ distant from the central radio galaxy, indicating that line surveys with small area would miss the bulk of CO sources in similar structures. 
Meanwhile, half of the CO emitters are found to have velocities larger than the escape velocities in assumption of virialization of the cluster core, which appears not gravitationally bound with the center mass. 
These unbound CO emitters extend to $4\times R_{200}$ but are barely found within the $R_{200}$ radius, which is consistent with the picture that the cluster core has been virialized and the outer regions are still in formation.

\spb Comparing to high redshift protoclusters known in the literature, this structure contains more CO emitters with relatively narrow line width and high luminosity being consistent with SMGs samples and indicating starburst/merger activities. 
The most starburst-like members have large sky distances to the center galaxy with $>0'.5-4'.0$, which is beyond the FoV of single pointings of VLA and ATCA. This indicates that small area observations could have severe bias in protocluster fields missing a bulk of starbursting members.

\spb We construct the CO luminosity function in this cluster, and find a high overdensity of luminous CO sources. The amplitude of the luminosity function, log($\Phi_*$/cMpc$^{-3}$~dex$^{-1}$)$=-2.16\pm0.49$, is 1.6$\pm$0.5 orders of magnitude higher than the CO luminosity density in blank fields, and also higher than the prediction from semi-analytical SHARK models by one order of magnitude. This indicates a high density of cold molecular gas $0.6-1.3\times10^{9}$ ${\rm M_{\odot}~cMpc^{-3}}$, significantly higher than probed for field galaxy samples by more than one order of magnitude. 
This gas density is also higher than predictions by semi-analytical SHARK models for protoclusters while it is comparable with the cold gas density in the hydrodynamical DIANOGA simulations. We attribute the underprediction of SHARK partially to the lack of very high density regions in their simulated box, $\sim(300 {\rm Mpc})^3$.

To summarize, this study shows that molecular line surveys in high-z protocluster are more efficient than surveys in blank fields,
and line surveys with large area and wide velocity range are of virtual importance to provide us an unbiased view of large structure and gas content in protocluster environment and thus testing our understanding of galaxy formation and numerical simulations.

\begin{acknowledgements}

The Australia Telescope Compact Array is part of the Australia Telescope National Facility which is funded by the Australian Government for operation as a National Facility managed by CSIRO. We acknowledge the Gomeroi people as the traditional owners of the Observatory site. This work is making use of observations made with ESO Telescopes at Chajnantor and Paranal under programme 084.A-1016(A), 083.F-0022, 088.A-0754(A) and 090.B-712(A). This research is based on observations made with the NASA/ESA Hubble Space Telescope obtained from the Space Telescope Science Institute, which is operated by the Association of Universities for Research in Astronomy, Inc., under NASA contract NAS 5–26555. These observations are associated with program 10327.
The authors thank Ian Smail, Stefano Borgani and Elena Rasia for helpful discussions in the preparation of this manuscript. 
The authors acknowledge B\"arbel Koribalski for help on ATCA observations and introduction of SoFiA. The authors thank Laura Lenki\'c and Alberto D. Bolatto for helpful discussions on CO luminosity function.
SJ, HD and JMR acknowledge financial support from the Spanish Ministry of Science, Innovation and Universities (MICIU) under grant AYA2017-84061-P, co-financed by FEDER (European Regional Development Funds).
HD acknowledges financial support
from the Spanish Ministry of Economy and Competitiveness
(MINECO) under the 2014 Ram\'on y Cajal program MINECO
RYC-2014-15686.
BE acknowledges official funding from the National Radio Astronomy Observatory. The National Radio Astronomy Observatory is a facility of the National Science Foundation operated under cooperative agreement by Associated Universities, Inc.
CL id funded by the ARC Centre of Excellence for All Sky Astrophysics in 3 Dimensions (ASTRO 3D), through project number CE170100013.
LB acknowledges ExaNeSt and Euro Exa projects, funded by the European Union Horizon 2020 research and innovation program under grant agreement No. 671553 and No. 754337, the agreement ASI-INAF n.2017-14-H.0; DIANOGA simulations have been carried out using MENDIETA Cluster from CCAD-UNC, which is part of SNCAD-MinCyT (Argentina); MARCONI at CINECA (Italy), with CPU time assigned through grants ISCRA B, and through INAF-CINECA and University of Trieste - CINECA agreements; at the Tianhe-2 platform of the Guangzhou Supercomputer Center by the support from the National Key Program for Science and Technology Research and Development (2017YFB0203300). The post-processing has been performed using the PICO HPC cluster at CINECA through an expression of interest. 
JBC thanks the National Science Foundation for support through grants AST-1714528 and AST-1814034, and additionally the University of Texas at Austin College of Natural Sciences.
CMC thanks the National Science Foundation for support through grants AST-1814034 and AST-2009577, the University of Texas at Austin College of Natural Sciences, and the Research Corporation for Science Advancement from a 2019 Cottrell Scholar Award sponsored by IF/THEN, an initiative of Lyda Hill Philanthropies.
\end{acknowledgements}

\bibliography{biblio}

\appendix
\label{sec:appendix}
\label{fig:appspectra}
\section{Gallery of CO(1-0) detections}

\begin{figure*}
\centering
\includegraphics[width=0.98\textwidth]{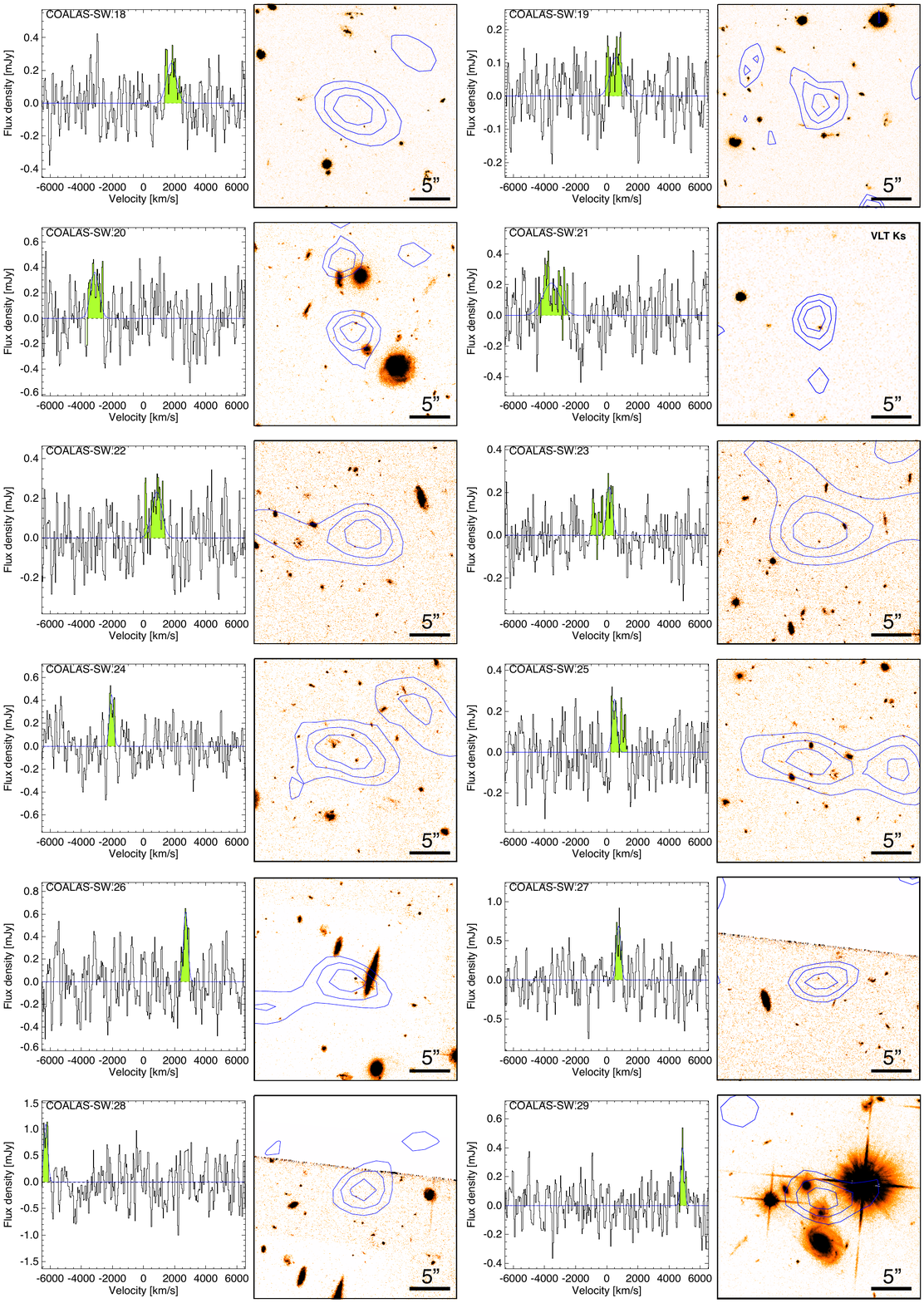}
\caption{
Spectra of SNR$=4$--5 detections in category A. This figure shares caption in Fig.~\ref{fig:spec_sn5_1}.
 \label{spec2}
}
\end{figure*}

\addtocounter{figure}{-1}
\begin{figure*}
\centering
\includegraphics[width=0.98\textwidth]{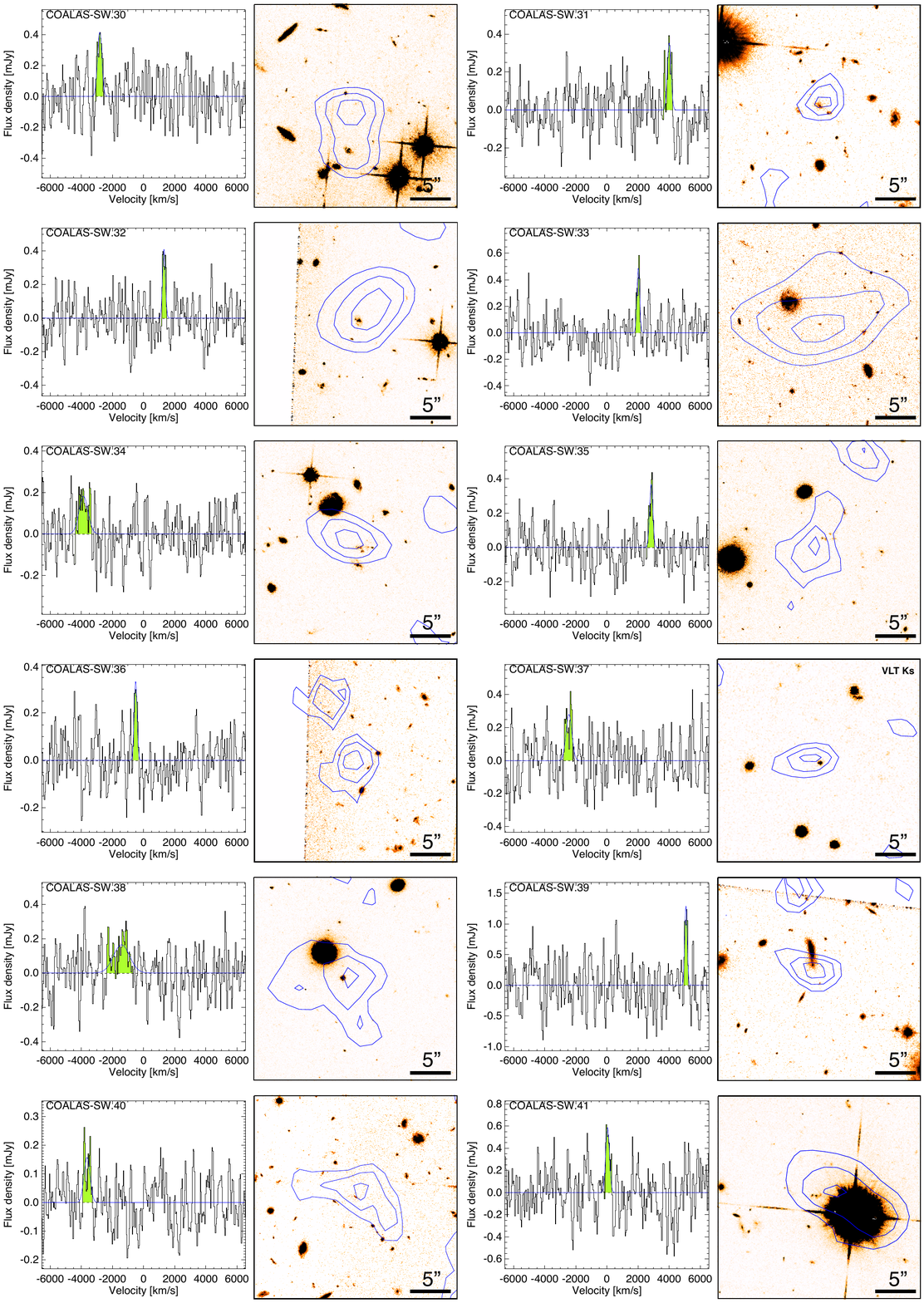}
\caption{
Continued.  
 \label{spec_sn45_1}
}
\end{figure*}

\addtocounter{figure}{-1}
\begin{figure*}
\centering
\includegraphics[width=0.98\textwidth]{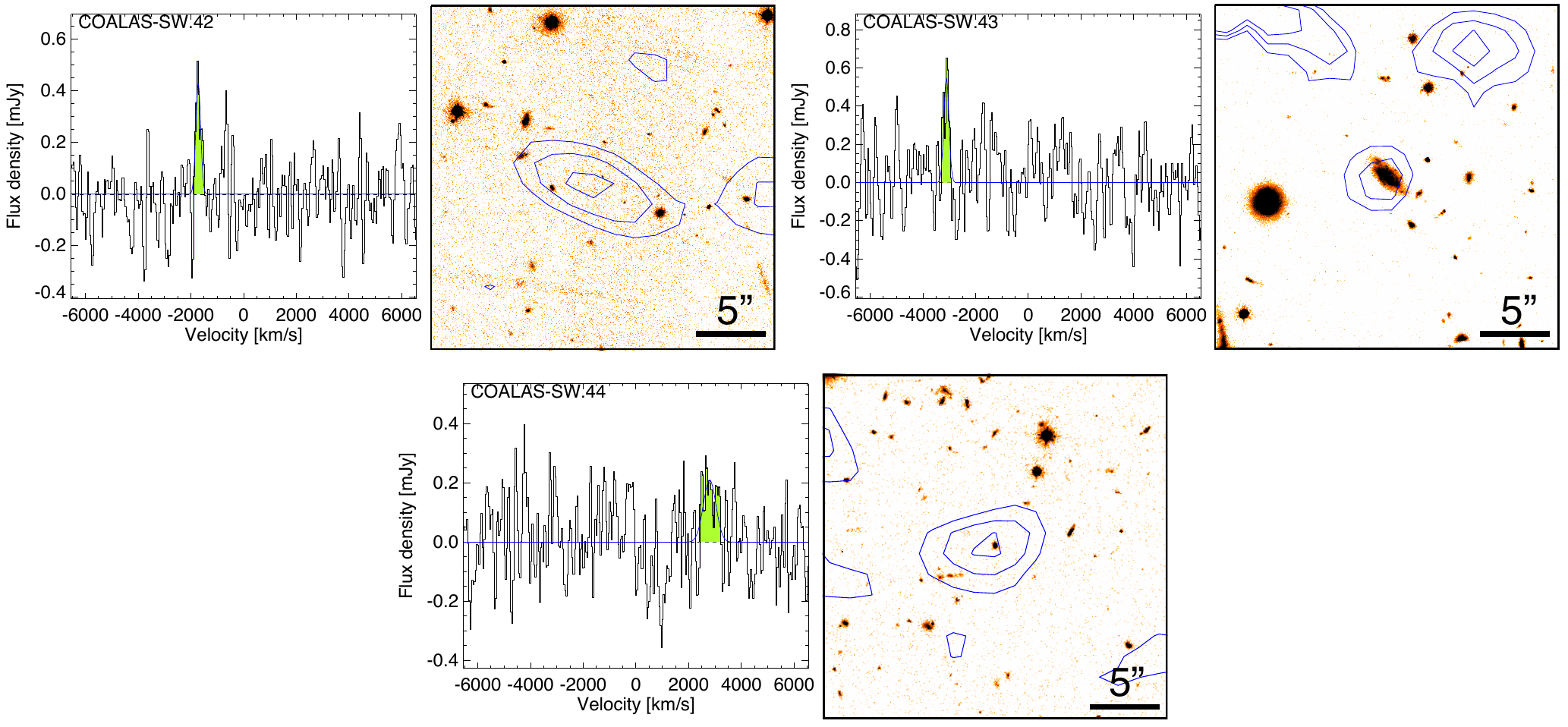}
\caption{
Continued.  
 \label{spec_sn45_2}
}
\end{figure*}

\begin{figure*}
\centering
\includegraphics[width=0.98\textwidth]{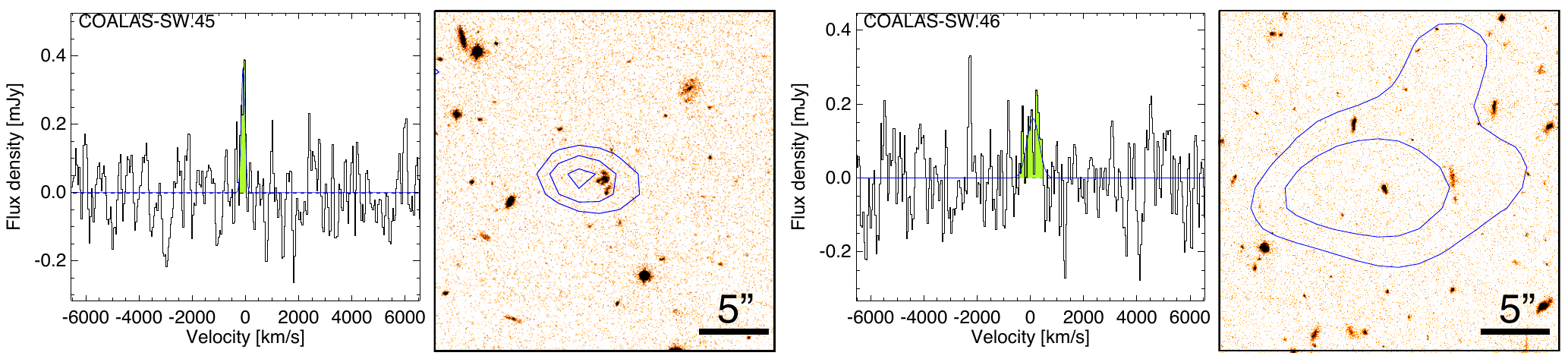}
\caption{
Spectra and intensity maps of category C detections. 
 \label{spec_catC}
}
\end{figure*}

\end{document}